\documentstyle[prc,aps,psfig]{revtex}
\def\Journal#1#2#3#4{{#1} {\bf #2}, #3 (#4)}

\def\NPA{{Nucl. Phys.} A}
\def\NPB{{Nucl. Phys.} B}
\def\PLB{{Phys. Lett.}  B}
\def\PR{{Phys. Rep.}}
\def\PRL{Phys. Rev. Lett.}
\def\PRC{{Phys. Rev.} C}
\def\PRD{{Phys. Rev.} D}

\def\ZPC{{Z. Phys.} C}

\def\P{{\mathbf p}}

\def\R{{\mathbf r}}

\def\a{\alpha}

\def\d{\delta}
\def\e{\epsilon}

\def\g{\gamma}

\def\p{\pi}
\def\q{\theta}

\def\n{\nu}

\def\s{\sigma}
\def\t{\tau}

\def\ce{{\cal E}}

\def\cm{{\cal M}}
\def\cp{{\cal P}}

\def\intps{\int \frac{d^3 \P}{(2\pi)^3}}

\def\exp{\mbox{\rm exp}}

\def\ra{\rightarrow}

\def\lra{\longrightarrow}
\def\llra{\longleftrightarrow}

\def\del{\mbox{$\partial$}}
\def\ncdot{\!\cdot\!}

\def\be{\begin{equation}}
\def\ee{\end{equation}}
\def\bea{\begin{eqnarray}}
\def\eea{\end{eqnarray}}
\def\eref#1{Eq.~(\ref{#1})}

\def\etrref#1#2#3{Eq.~(\ref{#1}), (\ref{#2}) and (\ref{#3})}

\def\fref#1{Fig.~\ref{#1}}

\newcommand{\ncom}{\newcommand}
\ncom{\vo}[1]{{\mathbf #1}}
\ncom{\vmo}[1]{{|\mathbf #1|}}
\ncom{\vt}[2]{({\mathbf #1}-{\mathbf #2})}
\ncom{\lan}{\langle}
\ncom{\ran}{\rangle}
\ncom\nonum{\nonumber \\}
\ncom\fx{}
\ncom\gsim{\mbox{\raisebox{-0.6ex}{\ $\stackrel {>}{\sim}$\ }}}
\ncom\lsim{\mbox{\raisebox{-0.6ex}{\ $\stackrel {<}{\sim}$\ }}}

\ncom{\half}{{1\over 2}}
\ncom{\third}{{1\over 3}}
\ncom{\fourth}{{1\over 4}}
\ncom{\fifth}{{1\over 5}}
\ncom{\sixth}{{1\over 6}}

\ncom\Tg{T_{eq\; g}}
\ncom\Tq{T_{eq\; q}}
\ncom\qg{\q_g}
\ncom\qq{\q_q}

\begin{document}
\draft
\preprint{LPTHE-Orsay 96/26, BI-TP 96/18}

\title{THERMAL AND CHEMICAL EQUILIBRATION\\
IN\\ RELATIVISTIC HEAVY ION COLLISIONS}

\author{S.M.H. Wong}

\address{
\footnote{Laboratoire associ\'e au Centre National
de la Recherche Scientifique}LPTHE, Universit\'e de Paris XI, 
B\^atiment 211, F-91405 Orsay, France 
\\
and 
\\
Fakult\"at f\"ur Physik, Universit\"at Bielefeld,
D-33501 Bielefeld, Germany}

\date{May 15, 1996}

\maketitle

\begin{abstract}

We investigate the thermalization and the chemical 
equilibration of a parton plasma created from Au+Au collision 
at LHC and RHIC energies starting from the early moment when the
particle momentum distributions in the central region become
for the first time isotropic due to longitudinal cooling. 
Using the relaxation time approximation for the collision 
terms in the Boltzmann equations for gluons and for quarks 
and the real collision terms constructed from the simplest 
QCD interactions, we show that the collision times have the 
right behaviour for equilibration. The magnitude of the quark 
(antiquark) collision time remains bigger than the gluon 
collision time throughout the lifetime of the plasma so 
that gluons are equilibrating faster than quarks both 
chemically and kinetically. That is we have a two-stage 
equilibration scenario as has been pointed out already by 
Shuryak sometimes ago. Full kinetic equilibration is 
however slow and chemical equilibration cannot be completed 
before the onset of the deconfinement phase transition assumed
to be at $T_c=200$ MeV.  By comparing the collision entropy 
density rates of the different processes, we show explicitly 
that inelastic processes, and \emph{not} elastic processes as 
is commonly assumed, are dominant in the equilibration of the 
plasma and that gluon branching leads the other processes 
in entropy generation. We also show that, within perturbative
QCD, processes with higher power in $\a_s$ need not be less 
important for the purpose of equilibration than those with 
lower power. The state of equilibration of the system has 
also a role to play. We compare our results with those of 
the parton cascade model.

\end{abstract}

%\pacs{PACS numbers: 12.38.Bx, 12.38.Mh, 24.85.+p, 25.75.+r}
\pacs{LPTHE-Orsay 96/26, \ BI-TP 96/18}

\section{Introduction}

A goal of the future heavy ion collision experiments at the 
relativistic heavy ion collider (RHIC) at Brookhaven and at 
the large hadron collider (LHC) at CERN is to find the 
quark-gluon plasma. The primary aim is of course to show 
that quarks and gluons can indeed be freed from their hadronic
``prison'' and exist as individual entities in a hot plasma.
Once this is realized, one can then turn to the diverse
physics of such a new state of matter. One of these is 
the relation of the various thermodynamic variables to
each other or in other words, the equation of state \cite{shury1}.
In order to probe this in experiments, an equilibrated quark-gluon
plasma is required. In this work, we look at how far can one
expect to have such a plasma in equilibrium. Because of the
importance of this question, various different approaches have
already been taken to address this issue. In particular, Shuryak
\cite{shury2} argued that equilibration of the plasma
proceeds via two stages in the ``hot gluon scenario''. First the 
equilibration of the gluons and then that of the quarks follows 
with a certain time delay. Thermal equilibration is quite short 
for gluon $\le$ 1 fm with high initial temperature of 440 MeV at 
LHC and 340 MeV at RHIC. However, these estimates are based
on thermal reaction rates for large and small angle scatterings
and on the assumption that one scattering is sufficient to 
achieve isotropy of momentum distribution. As has been shown
in \cite{heis&wang1} using a family of different power behaviours 
for the time-dependence of the collision time, the assumption of 
one scattering is sufficient is a serious underestimate. With 
a larger number of scatterings, using the same arguments as 
in \cite{shury2}, the initial temperature will be lowered and 
the thermalization time will be increased. Also, we argue that 
estimates based on using the scattering rate alone is incorrect, 
since in a medium, one must consider the difference of the 
scattering going forward and backward both weighed with 
suitable factors of particle distribution functions. Hence, the
process with the largest cross-section is not necessarily the
more important. However, we will show the two-stage 
equilibration scenario or in other words, gluons equilibrate
much faster than quarks and antiquarks. 

The other approach is the semi-classical parton cascade model
(PCM) \cite{geig&mull,geig,geig&kap}, which is based on solving a 
set of relativistic transport equations in full six-dimensional 
phase space using perturbative QCD calculation for the interactions,
predicts an equilibration time of 2.4 fm/c for Au+Au collision
at 200 GeV/nucleon. This approach, which uses a spatial and 
momentum distribution obtained from the measured nuclear structure 
functions for the partons as initial state, is very complicated. 
Due to the finite size of the colliding nuclei, it is hard
to clearly identify thermalization in terms of the expected 
time-dependent behaviours of the various collective variables 
\cite{geig}. But by fitting the total particle rapidity and 
transverse momentum distributions of the defined central volume, 
roughly identical temperatures are obtained \cite{geig} and hence 
the claim of thermalization. However, in terms of the same 
distributions of the individual parton components, this becomes 
less obvious to be the case \cite{geig&kap}. As was stated in
\cite{geig&kap}, the momentum distributions are not perfect 
exponentials and therefore there is no complete thermalization
in any case. 

We will look at this problem of equilibration using a much
simpler approach which is based on the Boltzmann equation and 
the relaxation time approximation for the collision terms. 
Initially used by Baym \cite{baym} to study thermal equilibration 
and has subsequently been used in the study 
of various related problems \cite{gavin,kaj&mat,heis&wang2,wong}.
The conclusion of these works is, in general, if the collision time
$\q$ which enters in the relaxation approximation, grows less fast
than the expansion time $\t$, then thermal equilibration can be
achieved eventually. In the case of the quark-gluon plasma, it is
not sufficient to know that equilibration will be achieved 
eventually because the plasma has not an infinite lifetime in 
which to equilibrate. We would like to know how far can it 
equilibrate before the phase transition. To answer such a 
question, we will use both the relaxation time approximation 
and the interactions obtained from perturbative QCD for 
the collision terms to determine $\q$. This approach has been
used previously to study both thermal and chemical equilibration
in a gluon plasma \cite{wong} where it was found that with the
initial conditions obtained from HIJING results, the gluon plasma 
had not quite enough time to completely equilibrate. In the 
present case of a quark and gluon parton plasma, quarks and gluons 
are treated as different particle species rather than as generic 
partons and so they have different time-dependent collision times.
As a result, they approach equilibrium at different rates and
towards different target temperatures. The latters will converge only 
at large times. It follows that the system can only equilibrate as one
single system at large times. This lends support to the two-stage 
equilibration scenario \cite{shury2}.

In an expanding system, particles are not in equilibrium early on
because interactions are not fast enough to maintain this so they
are most likely to start off free streaming in the beam direction
\cite{heis&wang2,dan&gyul}. Thermalization will be seen as the 
gradual reduction of this free streaming effect as interactions gain 
pace and momentum transfer processes are put into action to bring 
the particle momenta into an isotropic distribution. The present 
approach takes into account of these effects. 

As in the previous work \cite{wong}, isotropic momentaneously
thermalized initial conditions are used at both RHIC and LHC
energies. These are obtained from HIJING results after allowing
the partons to free stream until the momentum distribution 
becomes isotropic for the first time \cite{biro&etal1,lev&etal,wang}.
From then on, interactions are turned on but the distribution 
becomes anisotropic again due to the tendency of the particles 
to continue to free stream. It is the role of interactions to 
reduce this and to progressively bring the distributions into 
the equilibrium forms. We have shown that, surprisingly, 
kinetic equilibration in a pure gluon plasma is driven 
mainly by gluon multiplication and not gluon-gluon elastic 
scattering. In this paper, we include quarks and antiquarks
and consider the equilibration of a proper QCD plasma. We 
explicitly break down the equilibration process into each of 
its contributing elements and show which interactions are more 
important and hence uncover the dominant processes for 
equilibration. In fact, our result is {\em inelastic} 
interactions are most important for this purpose both for 
quarks and for gluons. 

Our paper is organized as follows. In Sect. \ref{sec:relax}, we 
describe the Boltzmann equations with the relaxation time approximation 
for two particle species. In Sect. \ref{sec:therm}, the time-dependent 
behaviour of the collision times, $\q$'s, necessary for equilibration 
will be analysed and extracted. The particle interactions entering 
into the collision terms and details of their calculations will be 
explained in Sect. \ref{sec:cal}. Initial conditions used will be 
given in Sect. \ref{sec:ic} and lastly the results of the evolution 
of the plasma will be shown and discussed in Sect. \ref{sec:result}.
We finish with a brief discussion of the differences with the
results of PCM.

\section{Relaxation Time Approximation for Two Particle Species}
\label{sec:relax}

In the absence of relativistic quantum transport theory derived from 
first principle of QCD
\cite{heinz1,elze&etal,elze&heinz,elze,heinz2,heinz3}, 
we base our approach on Boltzmann equation with both the
relaxation time approximation for the collision terms
and the real collision terms obtained from perturbative QCD. 
Treating quarks and gluons on different footings, we
write down the Boltzmann equations
\be {{\del f_i} \over {\del t}}+{\vo v}_{\P\; i} \ncdot 
    {{\del f_i} \over {\del \R}} = C_i (\P,\R,t)
\ee
where $f_i$ is the one-particle distribution and $C_i$ stands
for the collision terms and includes all the relevant 
interactions for particle species $i$ and $i=g, q,\bar q$.
Concentrating in the central region of 
the collision where we assumed to be spatially homogeneous, 
baryon free and boost invariant in the z-direction (beam direction) 
so that $f_q=f_{\bar q}$ and $f_i=f_i(\P_\perp,\P'_z,\t)$ where
$p'_z =\g (p_z-u p)$ with $\g=1/\sqrt {1-u^2}$ and $u=z/t$
is the boosted particle z-momentum component and $\t =\sqrt {t^2-z^2}$ 
is the proper time. Following Baym \cite{baym}, the Boltzmann equation
can be rewritten as
\be {{\del f_i} \over {\del \t}} \Big |_{p_z \t}
    =C_i(p_\perp,p_z,\t)
\label{eq:baymeq}
\ee
in the central region. Using the relaxation time approximation
\be C_i(p_\perp,p_z,\t)=- 
    {{f_i(p_\perp,p_z,\t)-f_{eq \; i}(p_\perp,p_z,\t)} 
     \over \q_i(\t)} \; 
\label{eq:relaxapp}
\ee
where $f_{eq \; i}$ is the equilibrium distribution
and $\q_i$ is the collision time for species $i$,
this allows us to write down a solution to \eref{eq:baymeq}.
\be f_i(\P,\t)=f_{0\; i}(p_\perp,p_z \t/\t_0) e^{-x_i}
      +\int^{x_i}_0 dx'_i e^{x'_i-x_i} 
      f_{eq\; i}(\sqrt{p^2_\perp+(p_z \t/\t')^2},T_{eq\; i}(\t')) \; ,
\label{eq:baymeqsol}
\ee
where 
\be  f_{0\; i}(p_\perp, p_z \t/\t_0) = \Big ( \exp 
     (\sqrt {p^2_\perp + (p_z \t/\t_0)^2}/T_0)/l_{0\; i}
     \mp 1 \Big )^{-1} \; ,
\ee
is the solution to \eref{eq:baymeq} when $C=0$ which is also the 
distribution function at the initial isotropic time $\t_0$,
with initial fugacities $l_{0\; i}$ and temperature $T_0$. 
It is of such a form because of the assumption of momentaneously
thermalized initial condition. The functions $x_i(\t)$'s, given by
\be  x_i(\t)=\int^\t_{\t_0} d\t'/ \q_i(\t')  \; ,
\ee
play the same role as $\q_i$'s in the sense that their 
time-dependent behaviours control thermalization. 
$T_{eq\; i}$, that appears in $f_{eq\; i}$, is the 
time-dependent momentaneous target equilibrium temperature
for the $i$ particle species. The two terms of equation 
\eref{eq:baymeqsol} can be thought of, up to exponential 
factor, as the free streaming (first term) and equilibrium 
term (second term). Whether species $i$ equilibrates or not 
depends on which of the two terms dominates. 

In the present case of two species, the energy conservation 
equations are, in terms of the equilibrium ideal gas energy densities
$\e_{eq\; g}=a_2 \Tg^4$, $\e_{eq\; q}=n_f b_2 \Tq^4$, $a_2=8 \p^2/15$,
$b_2=7\p^2/40$ and $n_f$ is the number of quark flavours,
\be {{d \e_i} \over {d \t}}+{{\e_i+p_{L\; i}} \over \t} 
    =-{{\e_i-\e_{eq\; i}} \over \q_i} 
\ee
and 
\be {{d \e_{tot}} \over {d \t}}+{{\e_{tot}+p_{L\; tot}} \over \t} =0
    \; ,
\ee
where $\e_{tot}=\sum_i \e_i$ and $p_{L\; tot}=\sum_i p_{L\; i}$,
or in other words
\be \sum_i {{\e_i-\e_{eq\; i}} \over \q_i} =0 \; .
\label{eq:e_cons}
\ee
The above equation only expresses the fact that energy loss of one 
species must be the gain of the other. The transport equations
of the different particle species are therefore coupled as they
should be. The longitudinal and transverse pressures are defined 
as before
\be p_{L,T\; i}(\t)=\n_i \intps {{p_{z,x}^2} \over p} 
    f_i (p_\perp,p_z,\t) \; ,
\label{eq:pres}
\ee
with $\n_g=2\times 8=16$ and $\n_q=2\times 3\times n_f=6\, n_f$, 
the multiplicities of gluons and quarks respectively.

Here the equilibrium target temperatures $\Tg$ and $\Tq$
cannot be the same in general since, as we will see in Sect. 
\ref{sec:result}, $\qg \ne \qq =\q_{\bar q}$. Therefore gluons
and quarks will approach equilibrium at different rates. Note
that energy conservation here {\em does not} mean
\be \e_g+2 \e_q = \e_{eq\; g} + 2 \e_{eq\; q}
\label{eq:eeqe}
\ee
since $\qg < \qq$ always, at least at small times, so gluon
energy density $\e_g$ will approach $\e_{eq\; g}$ faster than
$\e_q$ approaches $\e_{eq\; q}$ so the two equilibrium energy
densities should not be considered to be those which can coexist
at the same moment. This can only be true at large $\t$
when $\Tg \simeq \Tq$ and $\qg \simeq \qq$.
If \eref{eq:eeqe} were true, the condition for energy 
conservation \eref{eq:e_cons} could not hold when 
$\qg \neq \qq$. Since our QCD plasma is a dynamical 
system under one-dimensional expansion as well as particle 
production, the target temperatures $\Tg$ and $\Tq$ must be 
changing continuously and must approach each other at large times 
before the gluon and quark (antiquark) subsystems can merge 
into one system and exist at one single temperature. 
Likewise, we believe $\qg$ and $\qq$ should also converge 
to a single value at large times, unfortunately, this will 
take too long to happen in the evolution of our 
plasma although we can be sure that both $\qg$ and $\qq$ 
increase less fast than the expansion time $\t$ near the end 
of the evolution, a condition which, as has already been stated in
the introduction and we will see again in Sect. \ref{sec:therm},
is necessary for thermalization.

\section{Conditions on $\qg$ and $\qq$ for Thermalization}
\label{sec:therm}

Before considering the evolution of the QCD plasma 
under real interactions, we can deduce analytically, using 
\eref{eq:baymeq} and \eref{eq:baymeqsol}, the conditions
on the $\q_i$'s under which the plasma will come 
to kinetic equilibrium. Multiplying \eref{eq:baymeqsol} by
particle energy and integrating over momentum, we have the
equations for the $\e_i$'s. Further manipulating these gives,
\be \int^{x_i}_0 d x'_i \; e^{x'_i} \Big \{
    \t' h(\t'/\t) \Big (\e_{eq\; i}(\t')-\e_i(\t') \Big )
   -{d \over {d x'_i}} \Big (\t' h(\t'/\t) \e_i(\t') \Big ) 
    \Big \} =0 \; ,
\label{eq:cond_theta}
\ee
where 
\be h(r)=\int^1_0 dy \sqrt {1-y^2 (1-r^2)} 
    =\half \bigg (r+{\sin^{-1} {\sqrt {1-r^2}} \over {\sqrt {1-r^2}}}
     \bigg ) 
\ee
and $x'_i=x_i(\t')$. Supposing as $\t \ra \infty$, $x_g \ra \infty$ 
and $x_q \ra \infty$ then the integrand in \eref{eq:cond_theta} 
will be weighed by the $\t' \ra \infty$ or large $x'_i$ 
limit. It follows that the term within braces in \eref{eq:cond_theta} 
must be zero at large $\t'$ so using $h'(r)|_{r=1}=1/3$, we have
\be {{d \e_i} \over {d \t}}+{4 \over 3} {\e_i \over \t} 
    = -{{\e_i-\e_{eq\; i}} \over \q_i} \; .
\label{eq:hydro_i}
\ee
This means each species will undergo near hydrodynamic expansion 
at large $\t$ modified by energy lost to or energy gained from 
the other species. The latter should be small at such times.
Summing \eref{eq:hydro_i} over species, we obtain the energy 
conservation equation for a system undergoing hydrodynamic
expansion
\be {{d \e_{tot}} \over {d \t}}+{4 \over 3} {\e_{tot} \over \t} 
    = 0 \; ,
\label{eq:hydro}
\ee 
with $p_{L\; tot}=\e_{tot}/3$.

If one $\q_i$ is such that the corresponding 
$x_i \ra x_{i\; \infty} < \infty$ as $\t \ra \infty$
then hydrodynamic expansion does not apply to that species since
we have
\be {{d (\e_i \t)} \over {d \t}} =-{{(\e_i-\e_{eq\; i}) \t} \over \q_i} 
    -p_{L\; i} \; ,
\label{eq:x->x<inf}
\ee
where now $p_{L\; i} \neq \e_i/3$, so kinetic equilibrium is 
not established. The r.h.s. of \eref{eq:x->x<inf} is negative if
these particles are losing energy or gaining energy at a rate
less than $p_{L\; i}/\t$ at large $\t$. Therefore $\e_i \t$
must decrease towards a non-zero asymptotic value $(\e_i \t)_\infty$,
since $x_{i\; \infty}  < \infty \Longrightarrow \e_i \t >0$  
always, which results in a free streaming final state for 
these particles 
\be \e_i(\t \ra \infty) \sim (\e_i \t)_\infty /\t \; .
\label{eq:asymp_e}
\ee
A similar free streaming final state will be reached if the rate
of gaining energy is larger than $p_{L\; i}/\t$ at large $\t$.
In this case, although $\e_i \t$ is increasing, the $\e_j$ of
the other particle species with $x_j \ra \infty$ as 
$\t \ra \infty$ will be close to $\e_{eq\; j}$ and
so the energy transfer will be very small. One can deduce
that as $\t \ra \infty$
\be 1 \gg {{\e_{eq\; i}-\e_i} \over \q_i}  \ra 0
    \; > {p_{L\; i} \over \t} \; \Longrightarrow \;
    {{d (\e_i \t)} \over {d \t}} \ra 0 \; ,
\label{eq:onlot}
\ee
hence $\e_i \t\ra (\e_i \t)_\infty$. That is
$\e_i \t$ now increases towards some asymptotic value instead
of decreasing towards one as in the previous case.
But it ends up with a free streaming final state nevertheless. 
We do not consider the case where the relative rate 
$(\e_{eq\; i}-\e_i)\t/\q_i p_{L\; i}$ oscillates about one at
large $\t$ except to say that on the average 
$d (\e_i \t)/d \t \sim 0$ and so an average free streaming final 
state is likely. 

The last possibility where 
$x_i \ra x_{i\; \infty} <\infty$ as $\t \ra \infty$
for both particle species, \eref{eq:x->x<inf} applies to both.
Barring the case of the oscillating relative rate, one particle
species must lose energy and so by the above argument, a free 
streaming final state results. For the remaining particle species,
it does not matter whether $d (\e_i \t)/d\t$ is or is
not positive at large $\t$, these particles will also be in
a free streaming final state. If the rate is negative, then the 
same argument that leads to \eref{eq:asymp_e} applies. If it is 
positive, since the species that is losing energy is approaching 
free streaming so the energy transfer must go to zero. Then 
we are back to \eref{eq:onlot}.

The conclusions are therefore, depending on the time-dependent
behaviours of $\qg$ and $\qq$,
\begin{enumerate}

\item $x_g \ra \infty$ and $x_q \ra \infty$ as
$\t \ra \infty$ are required for the whole system to 
completely thermalize. 

\item $x_g \ra \infty$ and $x_q \ra x_{q\; \infty}
< \infty$ or $x_q \ra \infty$ and $x_g \ra 
x_{g\; \infty} < \infty$ as $\t \ra \infty$
imply that only the species with $x_i \ra \infty$ will 
thermalize, the other species will not equilibrate but free 
streams at the end. The system will end up somewhere between 
free streaming and hydrodynamic expansion.

\item Both $x_g \ra x_{g\; \infty} <\infty$ and 
$x_q \ra x_{q\; \infty} <\infty$ as $\t \ra \infty$
then the whole system will end up in a free streaming final state.

\end{enumerate}

One can understand these $x_i$ behaviours in terms of $\q_i$'s
by assuming simple power $\t$-dependence for the latters. 
One finds that $\q_i$'s must all grow slower than $\t$ for
the whole system to achieve thermalization. If either one
or more grow faster then a mixed or a complete free 
streaming final state results.

\section{Particle Interactions --- Collision Terms}
\label{sec:cal}

To investigate the evolution of a proper QCD plasma, we consider
the following simplest interactions at the tree level
\be gg \llra ggg \; \; , \; \; \; gg \llra gg \; , 
\label{eq:ggi}
\ee 
\be gg \llra q\bar q \; \; , \; \; \; g q \llra g q \; \; , 
    \; \; \; g\bar q \llra g\bar q \; ,
\label{eq:gqi}
\ee
\be q\bar q \llra q\bar q \; \; , \; \; \; qq \llra qq \; \;  , 
    \; \; \; \bar q \bar q \llra \bar q \bar q \; .
\label{eq:qqi}
\ee
As in \cite{biro&etal1,lev&etal,wang}, we include only the 
leading inelastic processes i.e. the first interaction of 
\eref{eq:ggi} and \eref{eq:gqi}\footnote{The first one of 
\eref{eq:qqi} could also be inelastic but here we give the same 
chemical potential to all the fermions so we do not consider
quark-antiquark annihilations into different flavours as
inelastic for our purpose.}. We will return to this point 
later on in Sect. \ref{sec:result}. 

In the solutions \eref{eq:baymeqsol} to the Boltzmann equations 
\eref{eq:baymeq}, there are two time-dependent unknown parameters 
$\q_i$ and $T_{eq\;i}$ for each species which very much control 
the particle distributions. To determine them, we need two 
equations each for gluons and for quarks. In order to show the 
relative importance of the various interactions 
\etrref{eq:ggi}{eq:gqi}{eq:qqi} in equilibration, we find 
these time-dependent parameters by constructing equations from 
the rates of energy density transfer between quarks (antiquarks) 
and gluons and the collision entropy density rates. 

From \etrref{eq:baymeq}{eq:relaxapp}{eq:baymeqsol},
the energy density transfer rates are
\be {{d \e_i} \over {d \t}}+{{\e_i + p_{L\; i}} \over \t} 
    = -{{\e_i-\e_{eq\; i}} \over \q_i} 
    = \n_i \intps \; p \; C_i (p_\perp,p_z,\t) = \ce_i \; ,
\label{eq:e_trans} 
\ee 
where $\ce_i$ is the energy gain or loss of species $i$ per unit
time per unit volume. As stated in Sect. \ref{sec:relax}, $\ce_i$'s 
must obey $\sum_i \ce_i =0$ for energy conservation.

The other equations, the collision entropy rates can be deduced
from the explicit expression of the entropy density in terms
of particle distribution function \cite{groot}
\be s_i(\t)=-\n_i \intps \Big \{f_i(\P,\t) \ln f_i(\P,\t)
          \mp (1 \pm f_i(\P,\t)) \ln (1 \pm f_i(\P,\t)) \Big \} \; ,
\ee
where the different signs are for bosons and fermions respectively.
They are, using again \etrref{eq:baymeq}{eq:relaxapp}{eq:baymeqsol}, 
\bea  \Big ( {{d s_i} \over {d \t}} \Big )_{coll} \fx & = & \fx
     -\n_i \intps \Big ( {{\del f_i} \over {\del \t}} \Big )_{coll}
     \ln \Big ({{f_i} \over {1 \pm f_i}} \Big )   \\
     \fx & = & \fx -\n_i \intps \; C_i (p_\perp,p_z,\t) 
     \ln \Big ({{f_i} \over {1 \pm f_i}} \Big )   
\label{eq:s_rate1}  \\
     \fx & = & \n_i \intps {{f_i -f_{eq\; i}} \over \q_i}
     \ln \Big ({{f_i} \over {1 \pm f_i}} \Big )  \; .
\label{eq:s_rate2} 
\eea

By using the explicit expression for the collision terms $C_i$'s 
constructed from the interactions \etrref{eq:ggi}{eq:gqi}{eq:qqi}
within perturbative QCD, \etrref{eq:e_trans}{eq:s_rate1}{eq:s_rate2}
allow us to solve for $\q_i$'s and $T_{eq\; i}$'s. 

The gluon multiplication contribution to $C_g$ is constructed from
the infrared regularized Bertsch and Gunion formula \cite{bert&gun} 
for the amplitude with partial incorporation of 
Landau-Pomeranchuk-Migdal suppression (LPM) for gluon emission and 
absorption \cite{biro&etal1,gyul&wang,gyul&etal,baier&etal} as 
in the previous work \cite{wong}. The explicit form of the 
gluon multiplication collision term and a discussion of the 
problem regarding how to incorporate the LPM effect correctly
can be found there also. The remaining binary interaction 
contributions to $C_i$ for particle $1$ is, as usual, given by
\bea C_{i\; 1}^{binary} \fx & = & \fx - \sum_{\cp_i} 
    {{S_{\cp_i} \n_2} \over {2 p_1^0}} \prod^4_{j=2} 
    {{d^3 \P_j} \over {(2\p)^3 2 p_j^0}} (2\p)^4 \d^4(p_1+p_2-p_3-p_4) 
    |\cm^{\cp_i}_{1+2 \ra 3+4}|^2    \nonum  
    & &  \times
    [f_1 f_2(1 \pm f_3)(1 \pm f_4) -f_3 f_4 (1 \pm f_1)(1 \pm f_2)]     
\eea
where the $\cp_i$ runs over all the binary processes in 
\etrref{eq:ggi}{eq:gqi}{eq:qqi} which involve species $i$, 
$|\cm^{\cp_i}|^2$ is the sum over final states and averaged over 
initial state squared matrix element, $S_{\cp_i}$ is a symmetry 
factor for any identical particles in the final states for the 
process $\cp_i$ and $\n_2$ is the multiplicity of particle 2.

We take $|\cm^{\cp_i}|^2 \; $'s from \cite{cut&siv} and infrared 
regularized them using either the Debye mass $m_D^2$ for gluons
or the quark medium mass $m_q^2$ for quarks to cut off any infrared 
divergence. These masses are now time-dependent quantities
in a non-equilibrium environment. With non-isotropic momentum
distribution, both the Debye mass \cite{biro&etal2,esk&etal} 
and the gluon medium mass, $m^2_g$, are directional dependent. This 
is, however, not the case for the quark medium mass, $m^2_q$, which 
remains directional independent as in equilibrium. The directional
dependence arises out of the cancellations between identical type
of distribution functions similar to those one finds in the 
derivation of hard thermal loops \cite{bra&pis,fre&tay}. 
To keep things simple, we removed the directional dependence from 
$m_D^2$ and use, for SU(N=3), to leading order in $\a_s$,
\be m^2_D (\t)=-8 \p \a_s \intps 
    {\del \over {\del |\P|}} \Big (N \, f_g +n_f \, f_q \Big )  
    \; .
\ee
For the quark medium mass, to the same order, we use
\be m_q^2 (\t)= 4\p \a_s \; \Big ({{N^2-1} \over {2\, N}}\Big )  
    \intps {1 \over {|\P|}} \; \big (f_g +f_q \big )  \; ,
\ee
which is just the equilibrium expression but with non-equilibrium
distribution functions. 

With these masses, we regularize the squared matrix elements by hand
and inserting the masses as follows. 
\bea |\cm_{gg \ra gg}|^2 \fx &=& \fx {{9\; g^2} \over 2} \; \bigg (
     3-{{u t} \over {(s+m_D^2)^2}} -{{u s} \over {(t-m_D^2)^2}}
     -{{s t} \over {(u-m_D^2)^2}} \bigg )
     \\
     |\cm_{gg \ra q \bar q}|^2 \fx &=& \fx {g^2 \over 6} \; \bigg (
     {t \over {(u-m_q^2)}} +{u \over {(t-m_q^2)}} \bigg ) 
     - {3 \over 8} \; {{u^2+t^2} \over {(s+4 m_q^2)^2}}   
     \\
     |\cm_{gq \ra gq}|^2 = |\cm_{g \bar q \ra g \bar q}|^2 
     \fx &=& \fx g^2 \bigg ( 1- {{2 u s} \over {(t-m_D^2)^2}} 
     -{4 \over 9} \; \bigg ( {u \over {(s+m_q^2)}}
     +{s \over {(u-m_q^2)}} \bigg ) \bigg )
     \\
     |\cm_{qq \ra qq}|^2 = |\cm_{\bar q \bar q \ra \bar q \bar q}|^2 
     \fx &=& \fx {{2\; g^2} \over 9} \; 
       \bigg ( {{2(s^2+t^2)} \over {(u-m_D^2)^2}}   
     + \d_{12} \; {{2(u^2+s^2)} \over {(t-m_D^2)^2}}  \nonum
     \fx & & \fx \; \; \; \; \; - \d_{12} \; 
     {4 \over 3} \; {s^2 \over {(t-m_D^2)(u-m_D^2)}} \bigg )   
     \\ 
     |\cm_{q \bar q \ra q \bar q}|^2 \fx &=& \fx {{2\; g^2} \over 9} \; 
     \bigg ( \d_{13} \d_{24} {{2(s^2+t^2)} \over {(u-m_D^2)^2}}   
     + \d_{12} \d_{34} \; {{2(t^2+u^2)} \over {(s+4 m_q^2)^2}} \nonum
     \fx & & \fx \; \; \; \; \; - \d_{12} \d_{13} \d_{34} \; 
     {4 \over 3} \; {t^2 \over {(u-m_D^2)(s+4 m_q^2)}} \bigg )  
\eea
where the $\d_{ij}$ signifies that the $i$ and $j$ quark or 
antiquark must be of the same flavour. This regularization amounts 
to screening spacelike and timelike infrared gluons by $m^2_D$ and 
$4 m^2_q$, respectively and infrared quarks by $m^2_q$. We stress
that this regularization is done in a very simple manner and with the
right order of magnitude for the cutoffs. Its aim is to get some 
estimates to the collision rates without involving too much with 
the exact and necessarily complicated momentum dependent form 
of the true infrared screening self-energies in an out-of-equilibrium 
plasma when their infrared screening effects should be in action. 
They should be the extension of the 2-point gluon and quark hard 
thermal loops \cite{bra&pis,fre&tay,weld,klim,tay&wong} to a 
non-thermalized environment. 

We should mention here that the choice of the pair of 
equations for solving the two time-dependent unknowns $\q_i$ 
and $T_{eq\; i}$ for each particle species is not unique.
One can equally use, for example, the rate equations for the
particle number density instead of the collision entropy 
density. With these other choices, the values of 
the different quantities are shifted somewhat due to the way 
that the initial conditions are extracted but there
is no qualitative different in the result. Our present choice 
has the distinct advantage that we can explicitly compare the
different processes using the collision entropy density rates. 
This will become clear when we show the results in
Sect. \ref{sec:result}.

\section{Initial Conditions}
\label{sec:ic}

To start the evolution, we use the same initial conditions for
the gluon plasma as before \cite{wong} based on HIJING result for 
Au+Au collision. The initial conditions for the quarks (antiquarks) 
are obtained by taking a ratio of $0.14$ for the number of initial
quark (antiquark) to the initial total number of partons as done in 
\cite{biro&etal1,lev&etal,wang}. The initial conditions are shown
in Table 1. One sees that the initial quark collision times 
are long compared to those of the gluons both at RHIC and LHC.
Especially at RHIC, the quark collision time is exceedingly long 
and so these particles are essentially free streaming initially. 
Taking these numbers as guides to how fast each particle
species is going to equilibrate, we can be sure already of a 
two-stage equilibration scenario \cite{shury2}. 

\begin{center}
\begin{tabular}{|c|c|c|} \hline\hline
\multicolumn{3}{|c|}{Initial Conditions} \\ \hline
\emph{\ } & \ \ \emph{RHIC}\ \ & \ \ \emph{LHC}\ \ \\ \hline
$\t_0$ (fm/c)  &  0.70  &  0.50  \\ 
$T_0$ (GeV) & 0.50 & 0.74 \\
$\e_{0\; g}$ (GeV/$\mbox{\rm fm}^3$) & 3.20   & 40.00 \\
$\e_{0\; q}$ (GeV/$\mbox{\rm fm}^3$) & 0.63   & 7.83 \\
$n_{0\; g} (\mbox{\rm fm}^{-3})$  & 2.15  & 18.00 \\
$n_{0\; q} (\mbox{\rm fm}^{-3})$  & 0.42  & 3.53 \\
$l_{0\; g}$  & 0.08 & 0.21 \\
$l_{0\; q}$ & 0.017 & 0.044 \\
$\q_{0\; g}$ (fm/c) & 2.18 & 0.73 \\ 
$\q_{0\; q}$ (fm/c) & 239.72  & 30.92  \\ 
\hline\hline
\end{tabular}
\end{center}

\begin{center}

TABLE 1. Initial conditions for the evolution of a QCD
plasma created in Au+Au collision at RHIC and at LHC 

\end{center}

\vspace{0.5cm}

Using the standard initial picture of heavy ion collisions as 
before, our evolution is started when the momentum distribution 
in the central region of the collision becomes, for 
the first time, isotropic due to longitudinal cooling.
The subsequent development is determined by the interactions
\etrref{eq:ggi}{eq:gqi}{eq:qqi}. In the case of a pure gluon plasma
\cite{wong}, it is clear that interactions bring the system towards
equilibrium and not towards some free streaming final state
which is a possible alternative as can be inferred from the 
analysis in Sect. \ref{sec:therm}. That is the interactions 
dominate over the expansion. In the present situation, we will see
that the same can certainly be said for the gluons and for the 
quarks at LHC but at RHIC, it is less clear for the latters. 
The equilibration time for quarks is at least several times 
longer than that of the gluons. 

Details for the procedure of the computation can be found in 
\cite{wong}. The values for the numerical parameters are the same
and in addition, we use $n_f=2.5$ to take into account of the reduced
phase space of strange quark. All time integrations are discretized 
and the rates are obtained at each time step necessary for forming 
the two pairs of equations \etrref{eq:e_trans}{eq:s_rate1}{eq:s_rate2}.  
One then solves the two equilibrium temperatures $\Tg$ and $\Tq$
from two 4th degree polynomials, one for each of the temperatures.
From these solutions, $\qg$ and $\qq$ are obtained and everything 
is then fed back into the equations for the next time step.

\section{Equilibration of the QCD Plasma}
\label{sec:result}

We show the results of our computation in this section. They
show clearly the collision times $\qg$ and $\qq$ hold the keys 
to equilibration as have been analysed in Sect. \ref{sec:therm}.
We will see shortly that as a result of the disparity between their
magnitudes at finite values of $\t$, the equilibration of quarks
and antiquarks lags behind that of the gluons both chemically
and kinetically. We will also identify the dominant processes
responsible for equilibration. They are {\em not} the commonly
assumed elastic scattering processes as already mentioned in
the introduction.  

When dealing with two particle species, one has several choices
as to when should the evolution be stopped. We choose to do this
when both the quark and the gluon temperature estimates drop
to 200 MeV. For gluons, this estimate is obtained by the near
equilibrium energy and number density expression 
\be \e_g = a_2 \, l_g \, T^4_g \mbox{\hskip 1cm and \hskip 1cm}
     n_g = a_1 \, l_g \, T^3_g \; ,
\ee
which are valid when the fugacity $l_g$ is near $1.0$ i.e. when 
the distribution functions can be approximated by 
$f_g(\P,\l_g,\t) = l_g f_g(\P,\l_g=1,\t)$. For quarks and antiquarks, 
we cannot do the same as $l_q$ has not time to rise above $0.5$ so 
instead, the temperature is estimated from the same quantities in
kinetic equilibrium but at small values of $l_q$
\be \e_q = 3 \, \n_q \, l_q \, T^4_q / \p^2 
    \mbox{\hskip 1cm and \hskip 1cm}
    n_q = \n_q \, l_q \, T^3_q / \p^2 \; .
\ee

\begin{figure}
\centerline{
\hbox{
{\psfig{figure=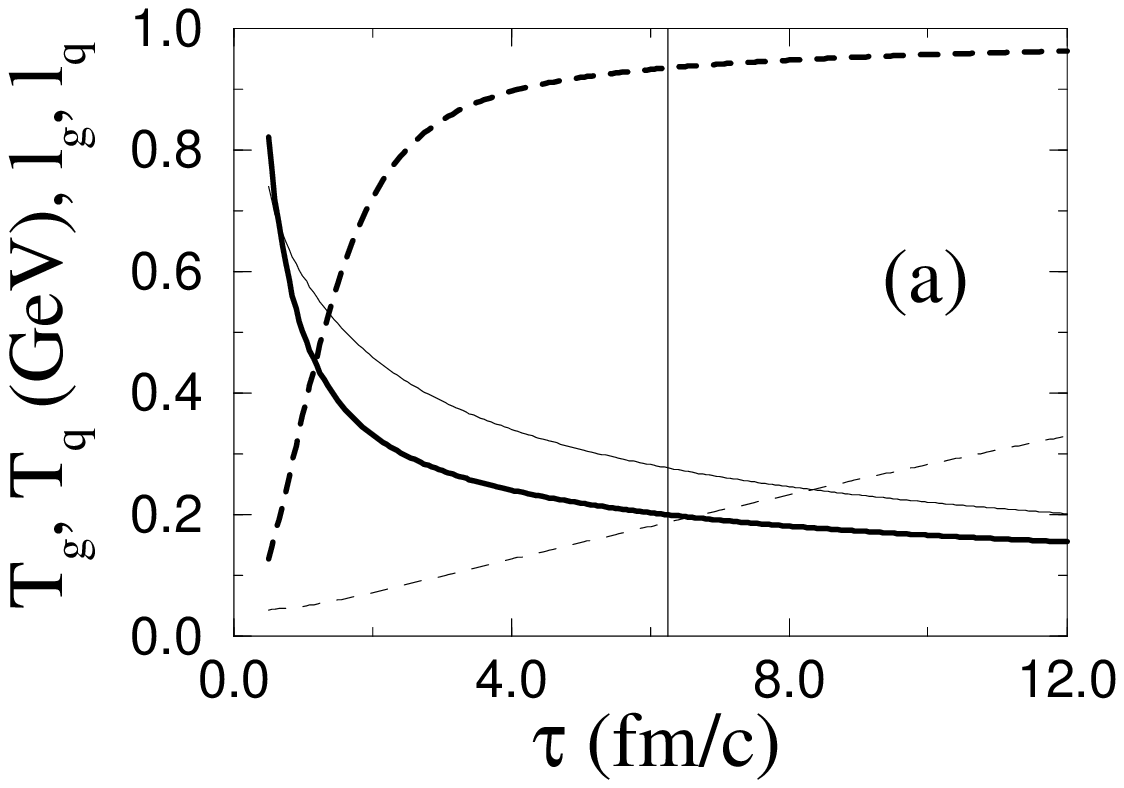,width=3.4in}} \ 
{\psfig{figure=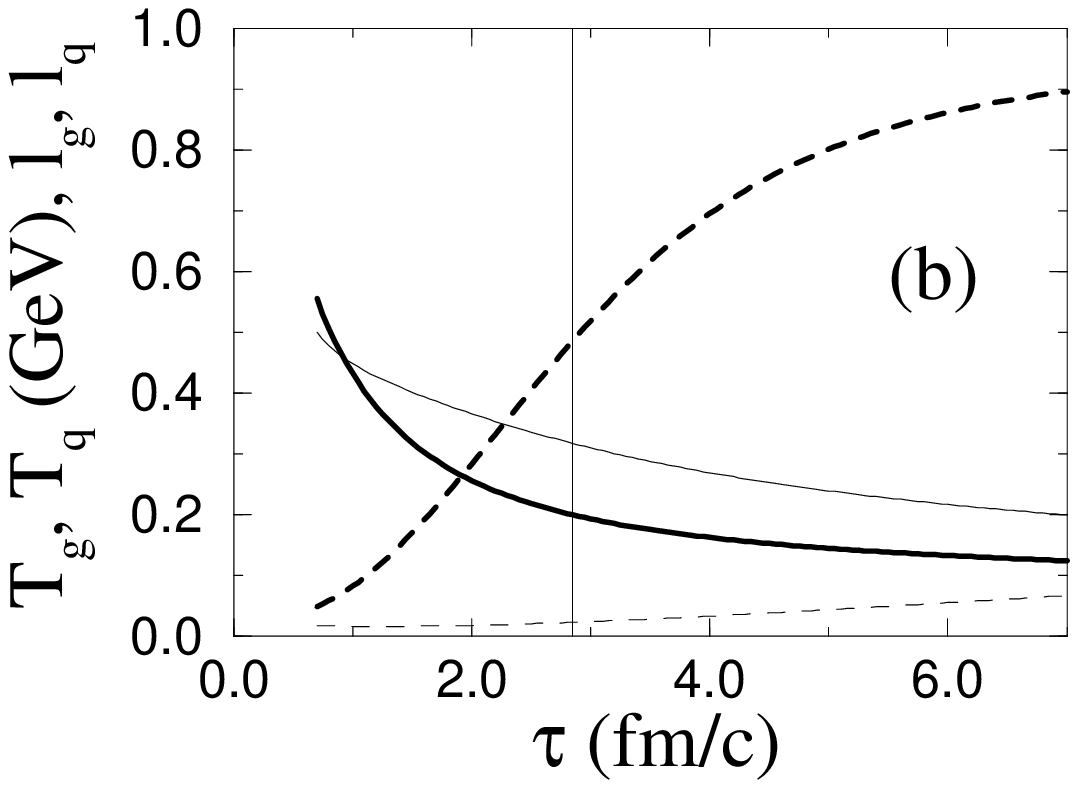,width=3.4in}}
}}
\caption{The time-dependence of the estimated temperatures for 
quarks and for gluons and their fugacities at (a) LHC and (b) RHIC.
The solid lines are the estimated temperatures $T_g$ (thick line)
and $T_q$. The dashed lines are the fugacities $l_g$ (thick line)
and $l_q$. Gluon chemical equilibration is much faster
than that of the quarks. The curves are stopped when all the
temperature estimates drop to 200 MeV. The vertical line indicates
when the gluon temperature reaches this value.}
\label{gr:T&l}
\end{figure}

\noindent 
These estimates are plotted in \fref{gr:T&l}. The vertical line
marks the point when the gluon temperature estimate (thick solid
line) drops to $200$ MeV. At this point, $\t \sim 6.25$ fm/c,
the fugacity (thick dashed line) is $l_g \sim 0.935$ at LHC and is 
$l_g \sim 0.487$ at $\t \sim 2.85$ fm/c at RHIC. On the same 
plots, the quark temperature (solid line) drops at a slower rate 
and the fermionic fugacity (dashed line) is also increasing much 
slower given the less favourable initial conditions and initially 
much slower quark-antiquark pair creation than gluon 
multiplication rate. In the end, the fermions are not too well 
chemically equilibrated and in fact, are still quite far away
from $1.0$. This is especially bad at RHIC. We note that comparing
to \cite{biro&etal1,lev&etal,wang}, in our case, gluons chemically 
equilibrate faster but quarks are slower. 

\begin{figure}
\centerline{
\hbox{
{\psfig{figure=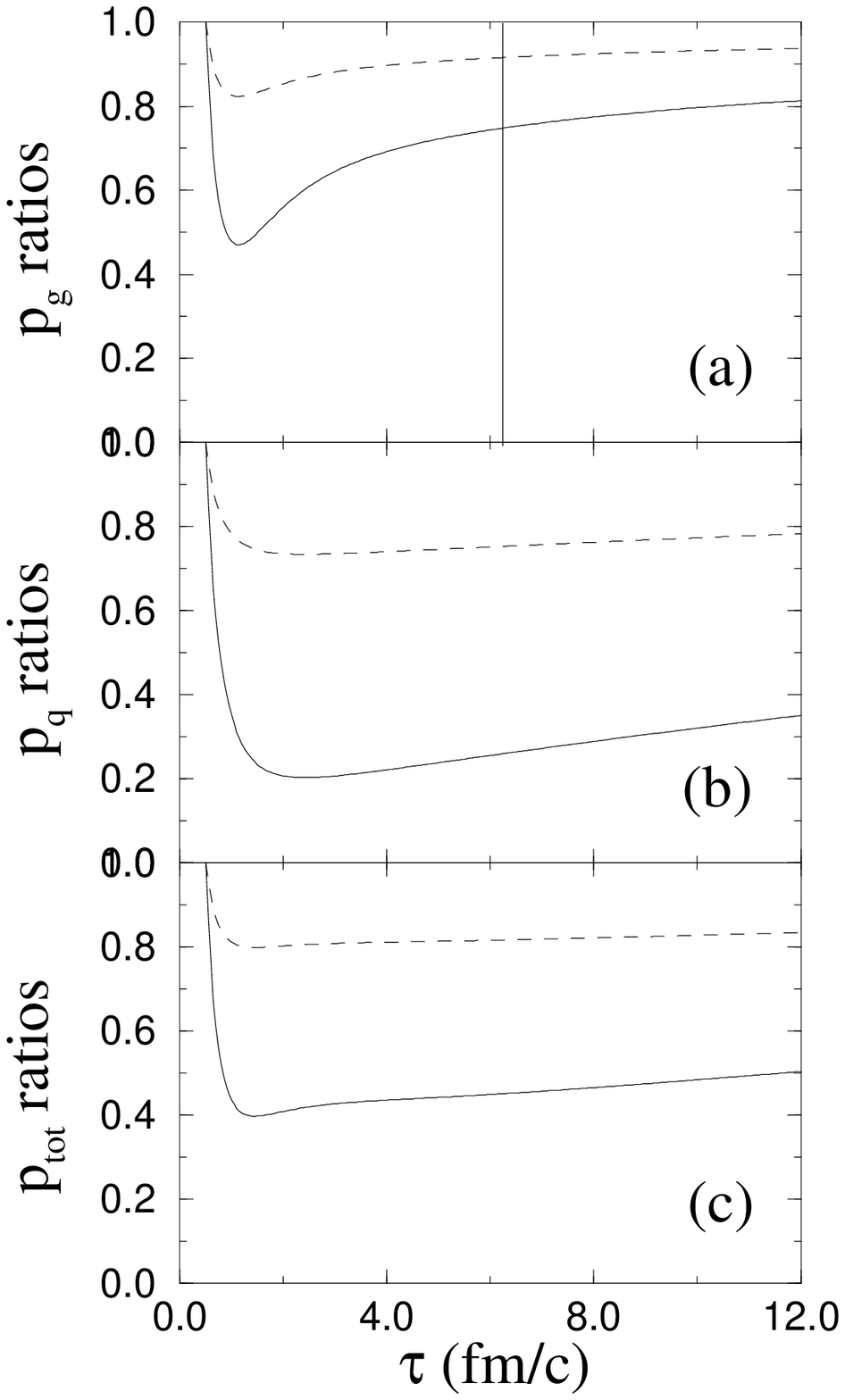,width=3in}} \ \
{\psfig{figure=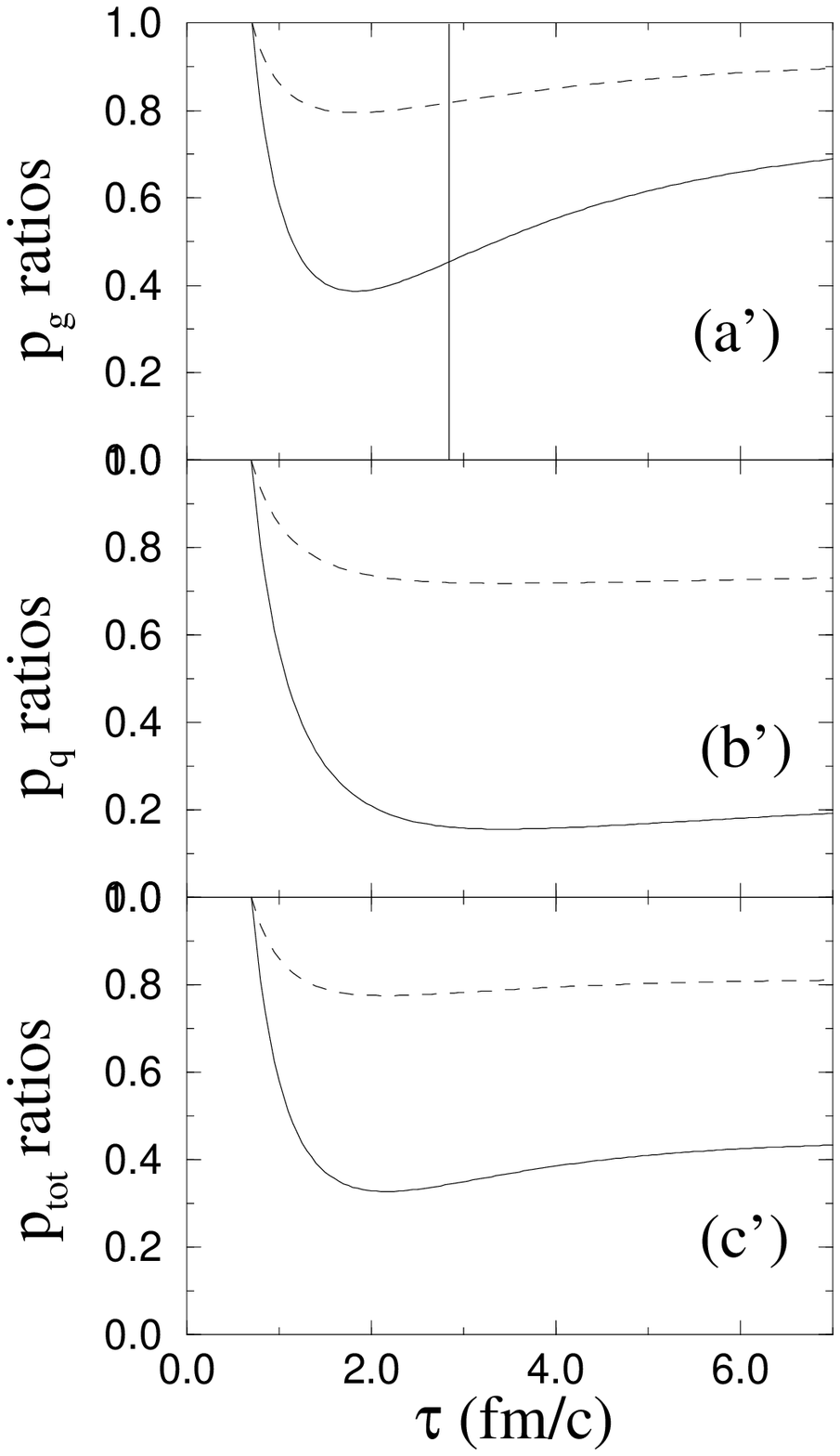,width=3in}}
}}
\caption{The ratios of the longitudinal pressure (solid line) 
and a third of the energy density (dashed line) to the transverse 
pressure, $p_L/p_T$ and $\e/3 p_T$ respectively for (a) gluons, 
(b) quarks and (c) the total sum at LHC. Graphs (a'), (b') 
and (c') are the same at RHIC.}
\label{gr:press}
\end{figure}

Unlike chemical equilibration, kinetic equilibration has no
simple indicators like the fugacities that can allow itself to be
simply quantified. One has to, instead, use the anisotropy of
momentum distribution as well as various reaction rates to
get an idea of the degree of kinetic equilibration. The 
former can be deduced from the ratios of the longitudinal pressure
and a third of the energy density to the transverse pressure,
$p_L/p_T$ and $\e/3 p_T$ respectively. Whereas from the elastic
scattering rates, one can deduce roughly how close the distribution
functions are to their equilibrium forms by virtue of the fact that
in local kinetic equilibrium, these rates are zero. The pressure
ratios $p_L/p_T$ (solid line) and $\e/3 p_T$ (dashed line) are
plotted in \fref{gr:press} (a) and (a') for gluons, (b) and (b') for 
quarks and (c) and (c') for the total sum. These ratios are indeed 
approaching $1.0$, the expected value after thermalization, but at 
different rates. Gluons are clearly equilibrating much faster than 
quarks which proceed rather slowly.

To show that these behaviours, although slow, are indeed the signs
of equilibration and that the plasma is not approaching
some free streaming final states, we can work out what their
behaviours should be in the latter case by taking the extreme
and let $\q_i \ra \infty$. From \eref{eq:pres}, as $\t \ra \infty$,

\vspace{2ex}
\hbox{\hspace{4.9cm} 
\raisebox{-2.5ex}{\vbox{
\hbox{
\(
    p_L \ra \p \, \t_0^3 \, \e_0 /4 \, \t^3   
\)
}
\hbox{
\(
    p_T \ra \p \, \t_0 \, \e_0 /8 \, \t   
\)
}
\hbox{
\(
    \; \; \e \: \ra  \p \, \t_0 \, \e_0 /4 \, \t  
\)
}
}}
\(
    \bigg \} \; \; \Longrightarrow \; \; \bigg \{ 
\)
\raisebox{-1ex}{\vbox{
\hbox{
\(
    p_L/p_T \ra 2 \, \t^2_0/ \t^2 \ra 0  
\)
}
\hbox{
\(
    \e/3 \: p_T \ra 2/3 
\)
}
}}
\vbox{\parbox{4cm}{
\be \; 
\ee
}}
}
\vspace{1ex}

\noindent 
where $\e_0$ is the initial energy density and the above ratios
are valid for both quarks and gluons in this extreme. Therefore
in the free streaming case, the first ratio should approach
zero and the second should approach 2/3. These are clearly not 
what we see in our plots. 

\begin{figure}
\centerline{
\hbox{
{\psfig{figure=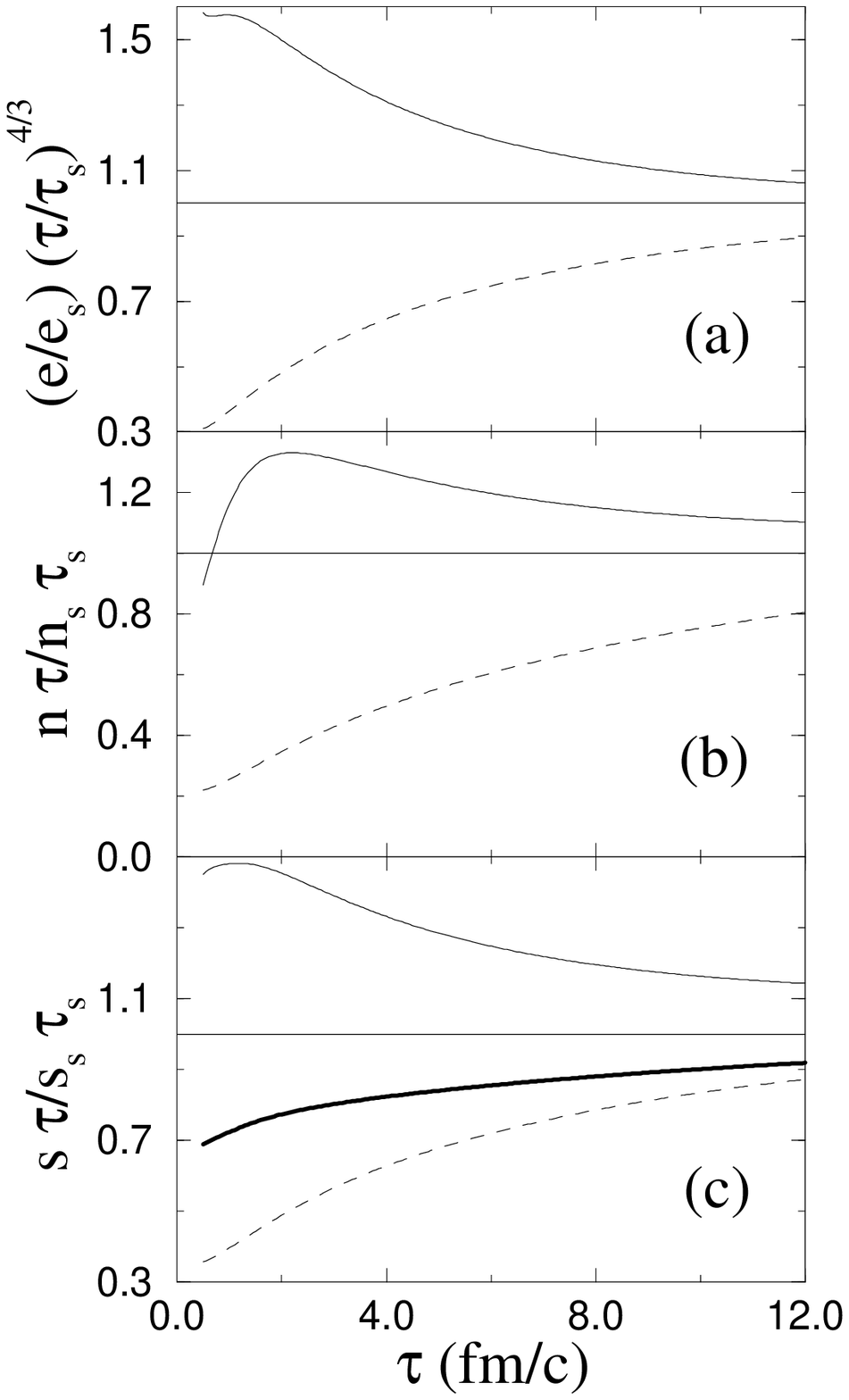,width=3in}} \ \
{\psfig{figure=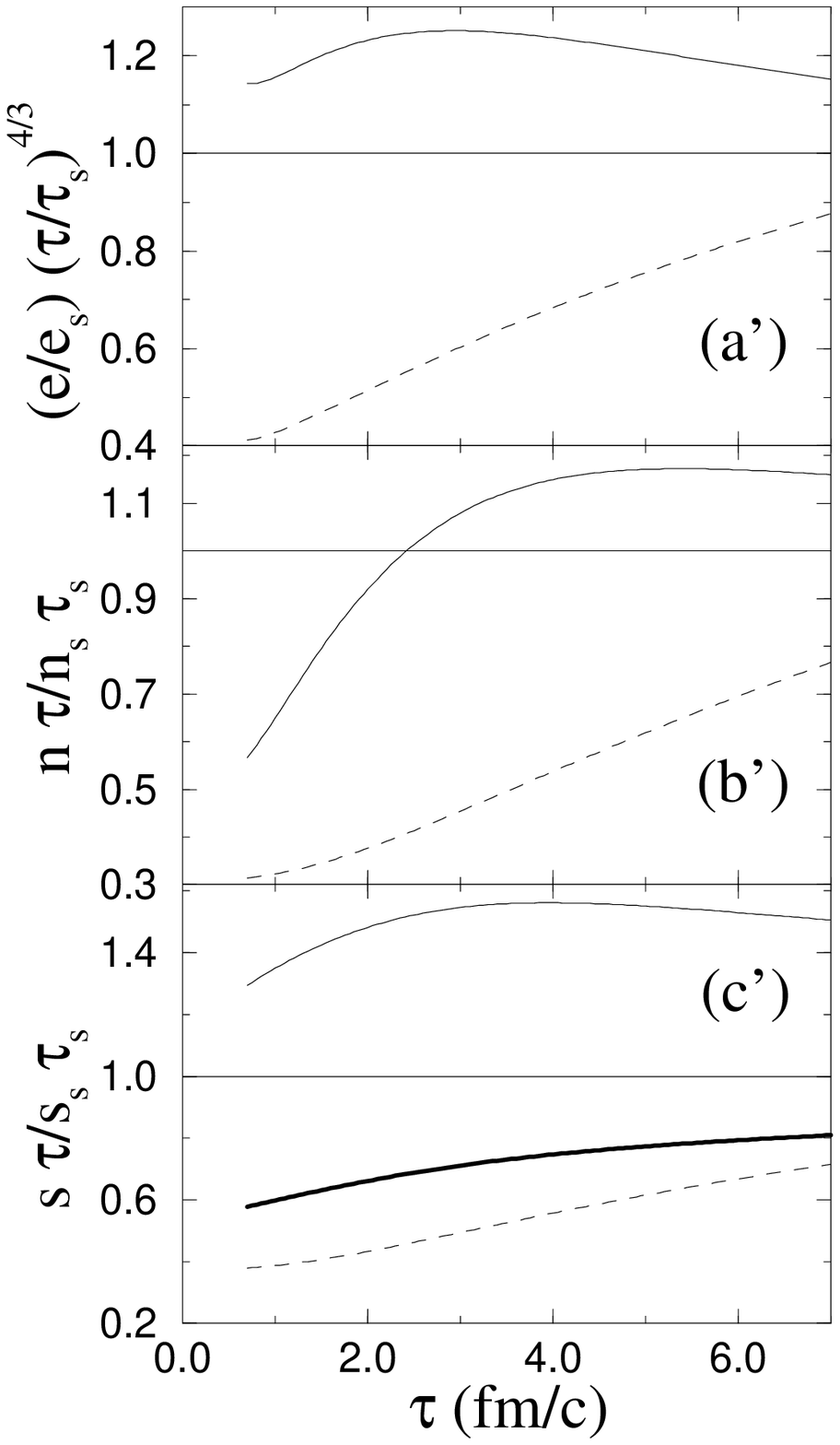,width=3in}}
}}
\caption{The scaled products of the collective variables
(a) energy density, (b) number density and (c) entropy density
and their expected inverse time-dependence in equilibrium 
$\t^{4/3}$, $\t$ and $\t$ respectively at LHC. Graphs (a'), (b') 
and (c') are the same at RHIC. The solid and dashed lines are 
for gluons and quarks respectively. The thick solid line 
in (c) and (c') is the scaled product of the total entropy
density and $\t$.} 
\label{gr:scaled}
\end{figure}

\begin{figure}
\centerline{
{\psfig{figure=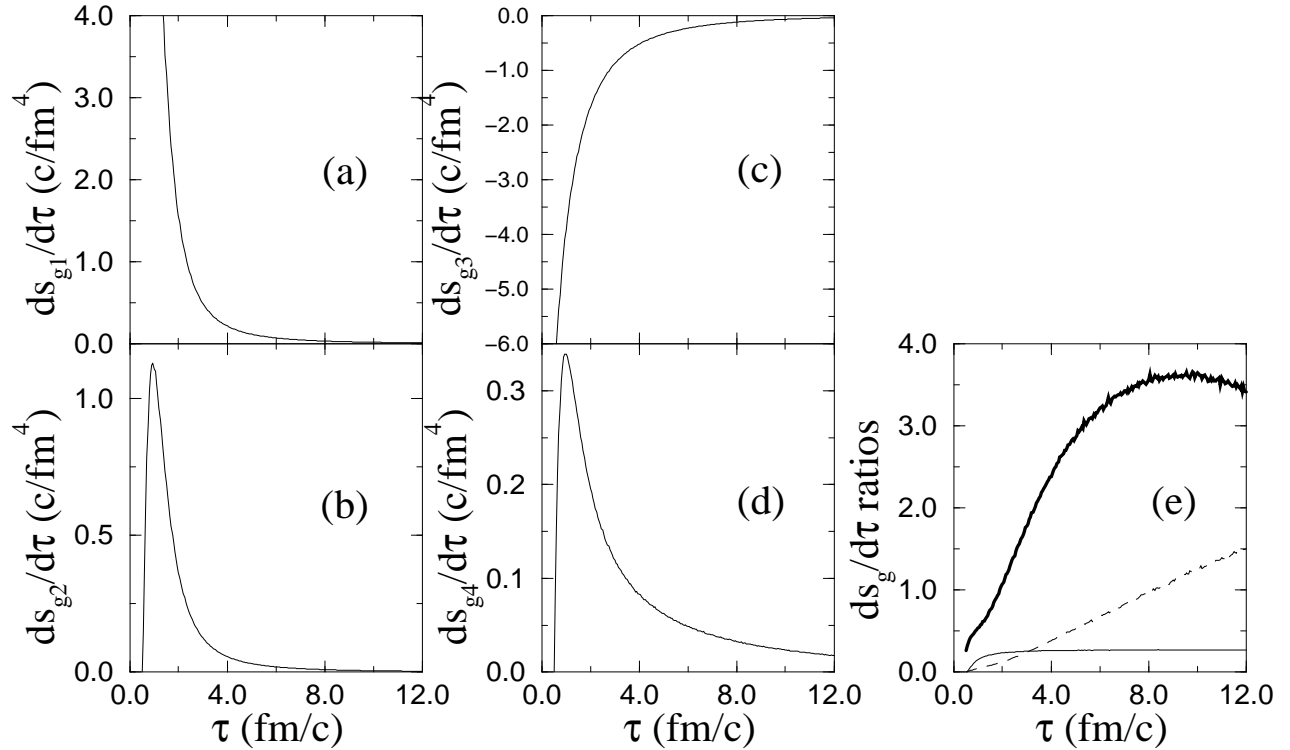,width=6.7in}}
}
\caption{ The time development of the different 
contributions to the total gluon collision entropy 
density rate at LHC. They are (a) $gg\llra ggg$, 
(b) $gg\llra gg$, (c) $gg\llra q\bar q$ and (d) $gq\llra gq$ 
or $g\bar q\llra g\bar q$. The curves of the elastic 
scattering processes in (b) and (d) have typical peaks at
maximum anisotropy in momentum distributions. The ratios
of the contribution (b) (thick line), (c) (solid line) and 
(d) (dashed line) to that of (a) are plotted in (e). This 
shows that first gluon multiplication dominates initially 
but is later overtaken by gluon annihilations into 
quark-antiquark pairs.}
\label{gr:s_g-lhc}
\end{figure}

\begin{figure}
\centerline{
{\psfig{figure=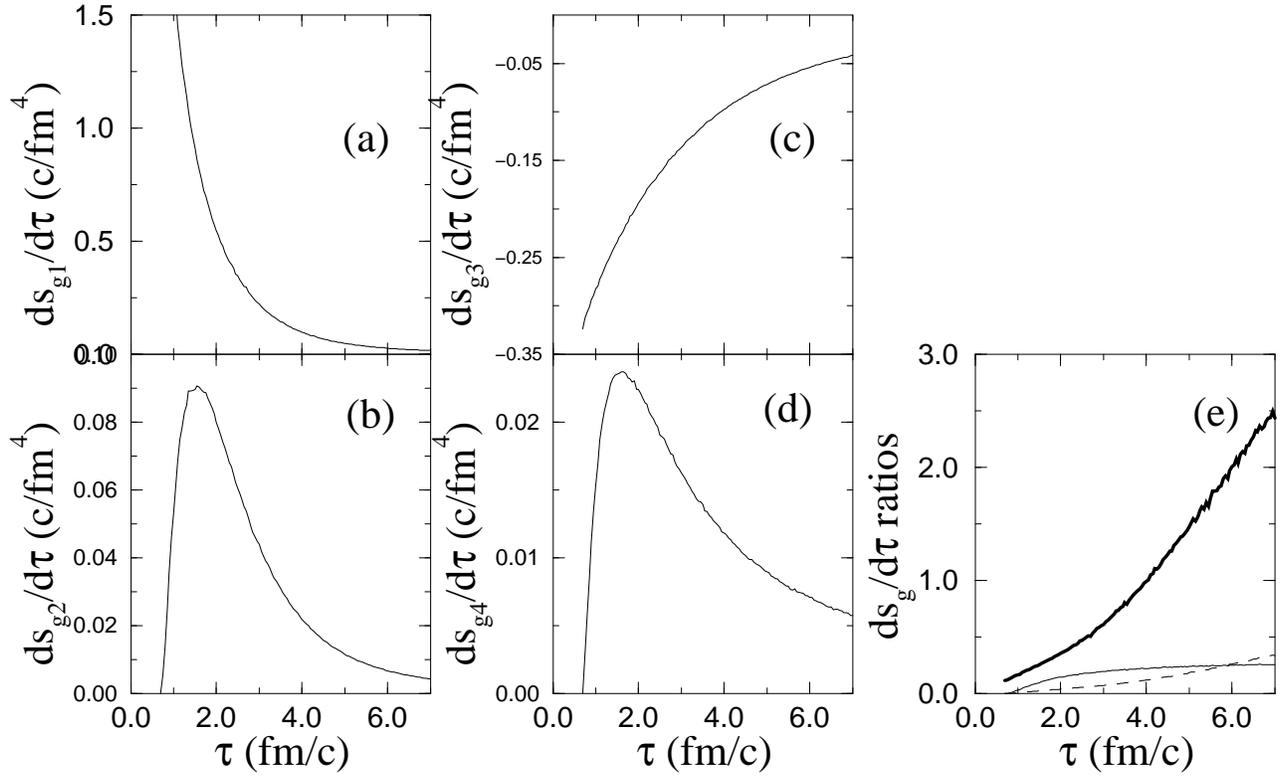,width=6.7in}}
}
\caption{The time development of the same contributions 
to the total gluon collision entropy density rate 
as in \fref{gr:s_g-lhc} but at RHIC. The same ratios
between the different contributions as at LHC are plotted 
in (e).}
\label{gr:s_g-rhic}
\end{figure}

\begin{figure}
\centerline{
{\psfig{figure=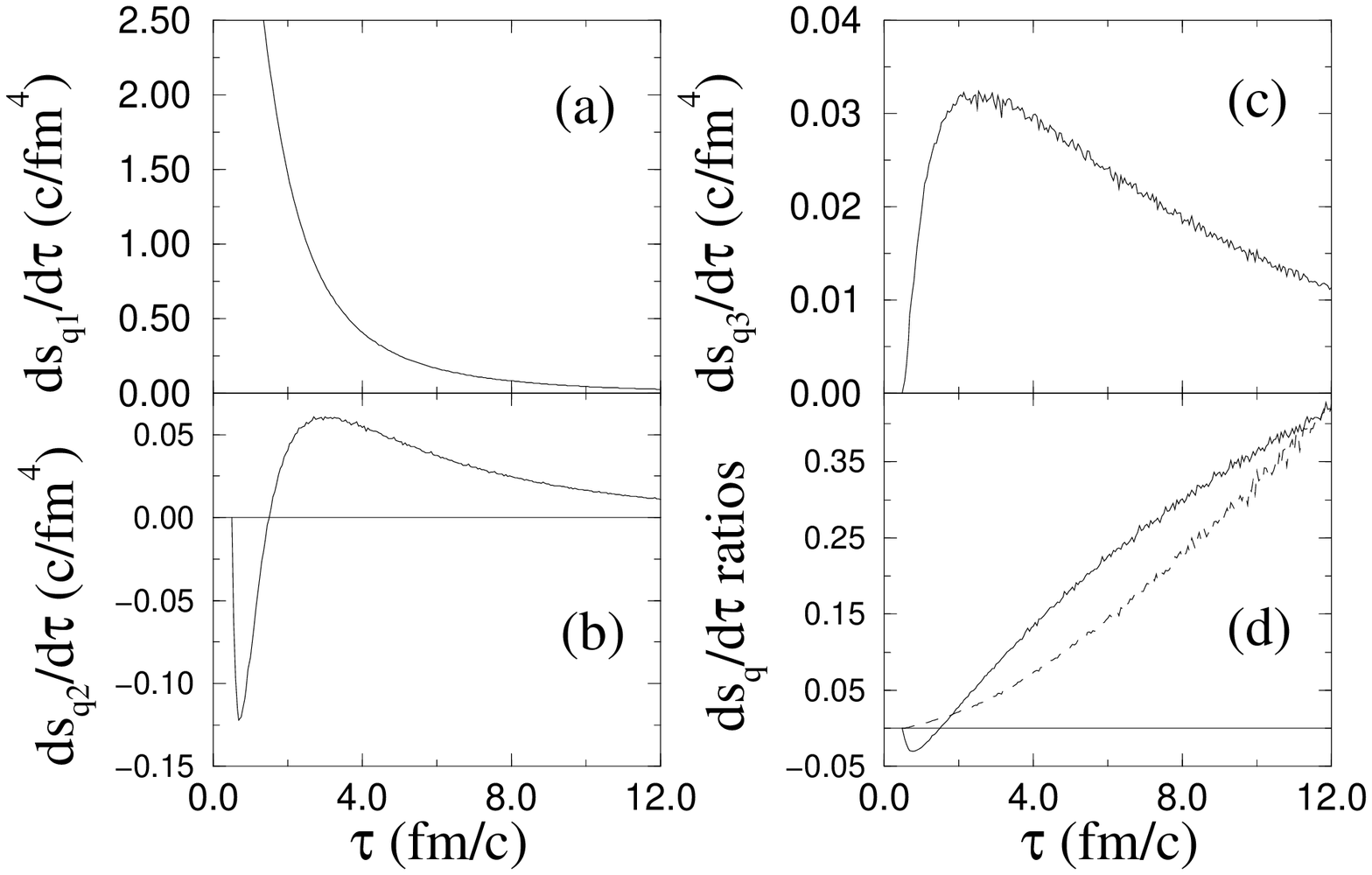,width=5in}}
}
\caption{ The time development of the different 
contributions to the total quark collision entropy density 
rate at LHC. They are (a) $gg\llra q\bar q$ 
(b) $gq\llra gq$ or $g\bar q\llra g\bar q$ and
(c) the sum of the contributions of all fermion elastic 
scattering processes $qq\llra qq$, $q\bar q\llra q\bar q$ and 
$\bar q\bar q\llra \bar q\bar q$. The ratios of the contribution
(b) (solid line), (c) (dashed line) to that of (a) is plotted 
in (d). This shows that throughout the lifetime of the 
QCD plasma, gluon annihilations into quark-antiquark 
pairs dominates in the equilibration of the fermions.}
\label{gr:s_q-lhc}
\end{figure}

\begin{figure}
\centerline{
{\psfig{figure=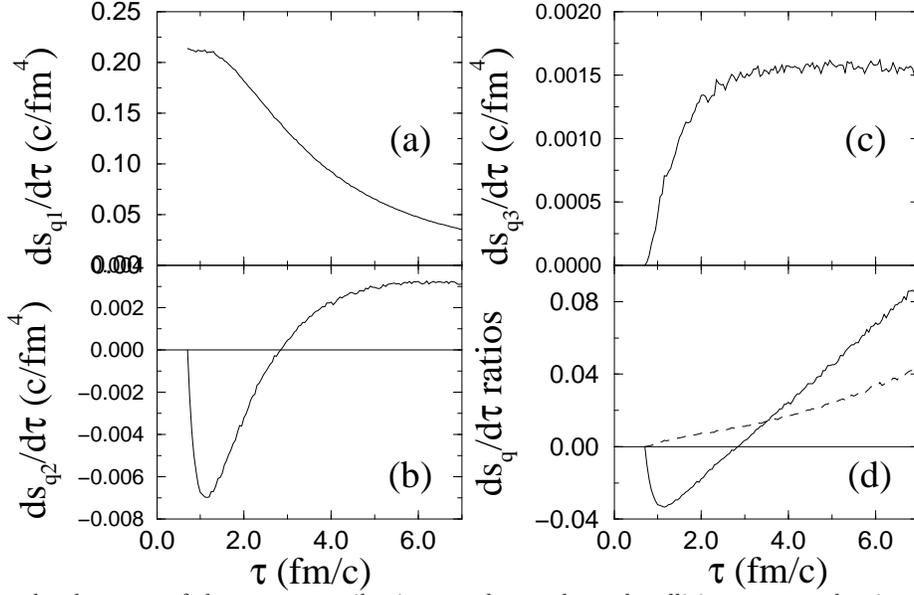,width=5in}}
}
\caption{The time development of the same contributions 
to the total quark collision entropy density
rate as in \fref{gr:s_q-lhc} but at RHIC. The same ratios
as at LHC are plotted in (d). They show again inelastic
process dominates.}
\label{gr:s_q-rhic}
\end{figure}

To best get an idea of how close the distribution functions are
to the equilibrium forms, the $gg$ and $qq$ or $\bar q\bar q$ 
elastic scattering processes are ideal for this. These are shown 
in \fref{gr:s_g-lhc} and \fref{gr:s_g-rhic} (b) for gluon and 
\fref{gr:s_q-lhc} and \fref{gr:s_q-rhic} (c) for
quark. Note that the peaks of these collision entropy rates 
coincide with the corresponding mininum points of the pressure 
ratios. As expected, the rates maximize at maximum anisotropy 
in momentum distribution. They all rise rapidly from zero at $\t_0$ 
when the interactions are turned on. The subsequent return to zero 
or the approach of the distribution functions to their equilibrium 
forms are, however, much less rapid. They only do so 
progressively as can be deduced already from the pressure ratio 
plots. 

Having shown chemical and kinetic equilibrations separately, we 
present now the actual approach of the collective variables towards 
the equilibrium values. Since we are more interested in the 
behaviour of their time-dependence than their absolute magnitudes,
we multiplied them by their expected time-dependence and 
scaled these by taking a guess at the corresponding asymptotic
values from the tendency of the curves. The results are plotted
in \fref{gr:scaled}. They are 
$\e_i \t^{4/3}/\e_{s\, i} \t_{s\, i}^{4/3}$, 
$n_i \t/n_{s\, i} \t_{s\, i}$ and $s_i \t/s_{s\, i} \t_{s\, i}$
in the figures (a) and (a'), (b) and (b') and (c) and (c') 
respectively. All these should be nearly constant 
with respect to time at large $\t$. The solid lines 
are for gluons and the dashed ones are for quarks. 
They showed that the curves do behave in such a way
for the eventual constant behaviour. This feature
is much clearer at LHC than at RHIC which only 
reconfirms the previously deduced result of faster equilibration 
at LHC than at RHIC. Note that for gluons, the quantities are
approaching the corresponding asymptotic values from above,
whereas for quarks, this approach is from below. This is 
because of the simple reason that there is a net conversion
of gluons into quark-antiquark pairs via $gg \llra q\bar q$.
The corresponding collision entropy density rate is negative as
shown in \fref{gr:s_g-lhc} and \fref{gr:s_g-rhic} (c). 
We will see that this same interaction becomes dominant 
in the later part of the evolution
later on when we compare the importance of the different 
processes. So gluons are losing energy, number and 
entropy to the fermions. This has to be so before the 
system as a whole can settle into complete equilibrium. 
The thick solid lines in \fref{gr:scaled} (c) and (c')
show the scaled total entropy per unit area in the central 
region which give an idea of the state of the system as 
a whole. They show that although the entropy of the individual 
subsystem can decrease, the total value must increase in 
accordance with the second law of thermodynamics. 

\begin{figure}
\centerline{
\hbox{
{\psfig{figure=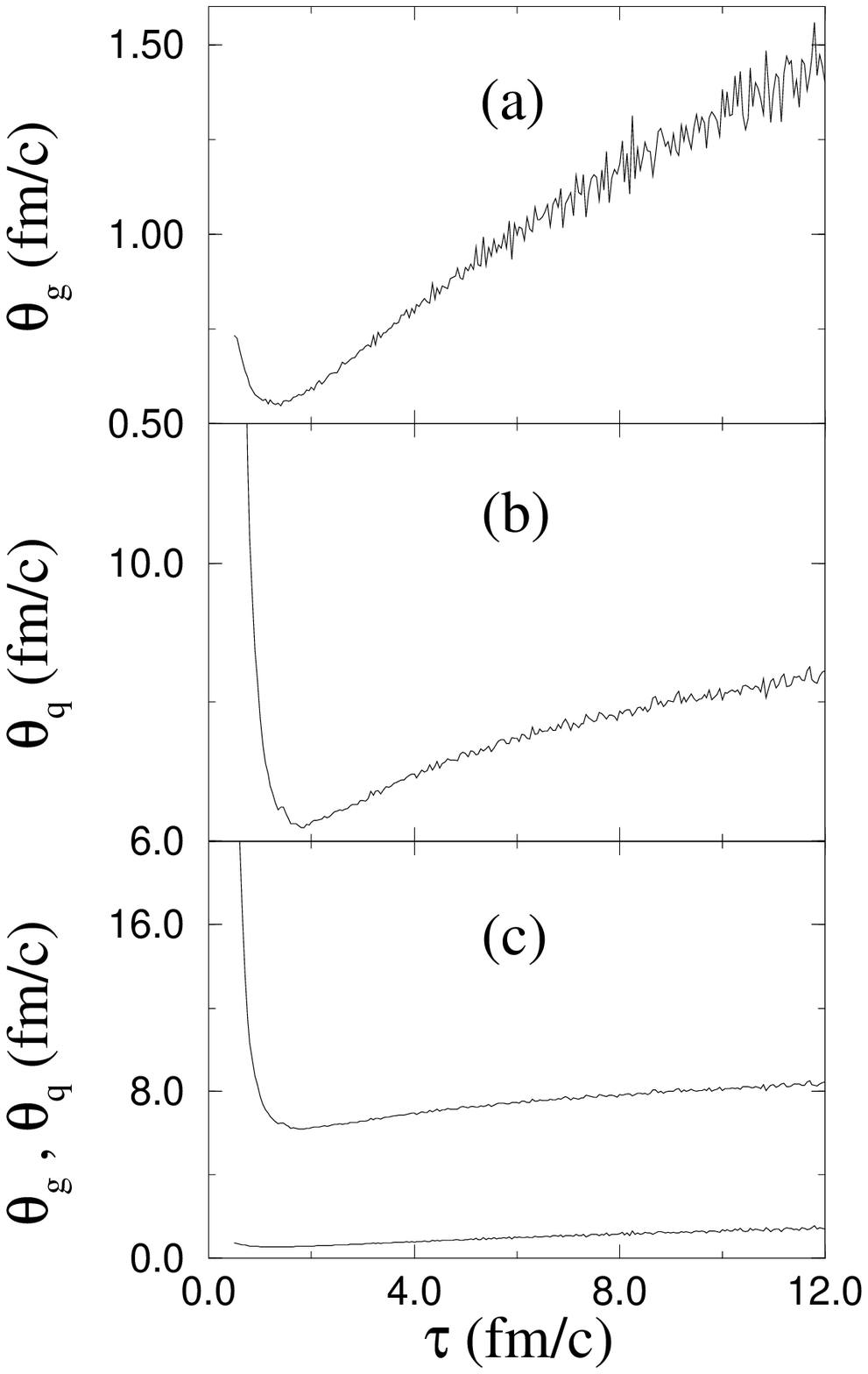,width=3in}} \ \
{\psfig{figure=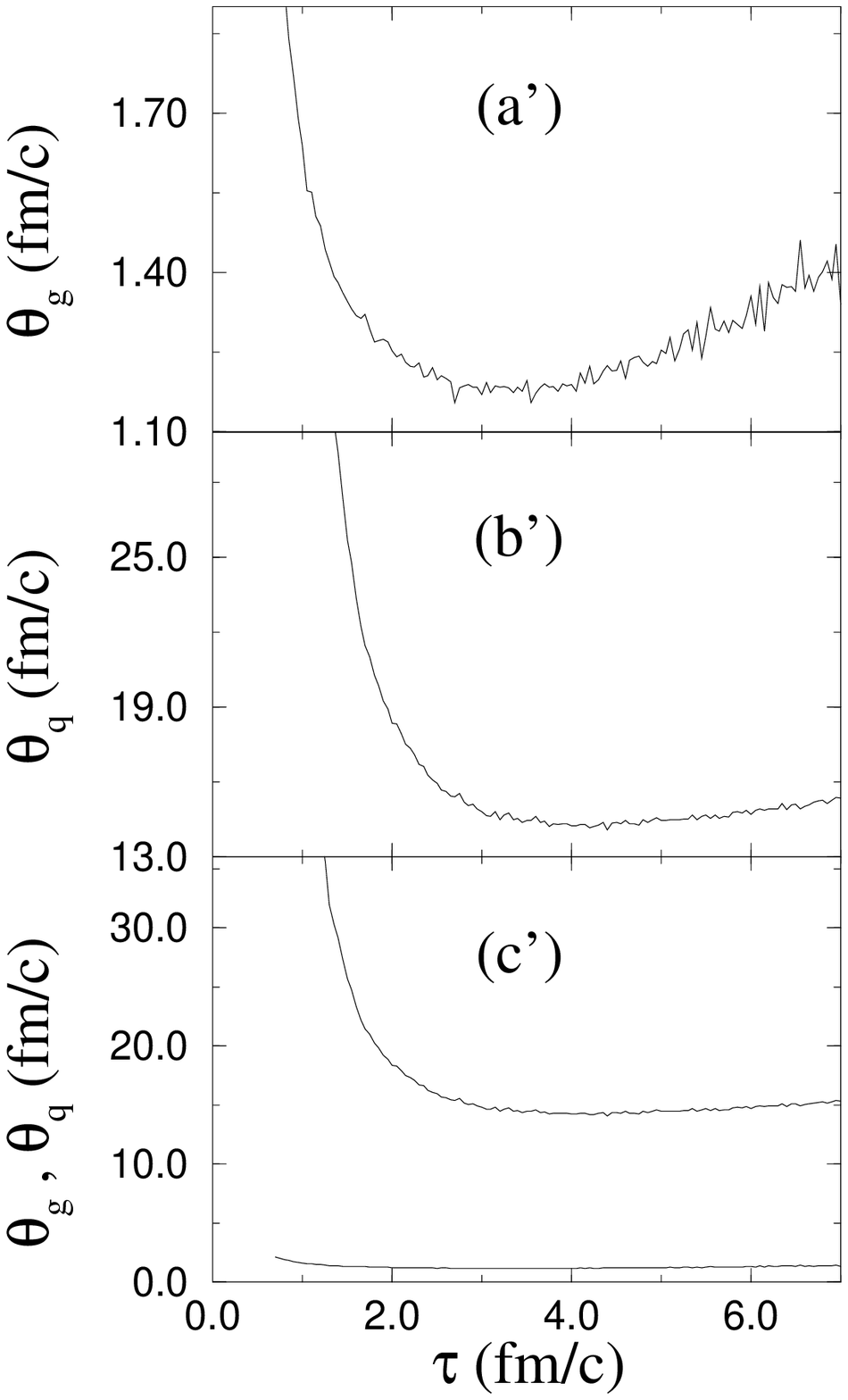,width=3in}}
}}
\caption{The time-dependence of the collision time
(a) for gluons $\qg$ and (b) for quarks $\qq$ at LHC. 
Their values are compared in (c). $\t$ overtakes first 
$\qg$ and later $\qq$ also. Graphs (a'), (b') and (c')
are the same at RHIC. In this case, $\t$ only has time 
to overtake $\qg$ but not $\qq$.}
\label{gr:relax_t}
\end{figure}

The figures discussed above show that the plasma 
is indeed approaching equilibrium and that interactions are 
fast enough to dominate over the Bjorken type one-dimensional 
scaling expansion. 

As we analysed in Sect. \ref{sec:therm}, thermalization is governed
by the $\q_i$'s. How fast this will proceed depends on their magnitudes 
and what is the actual final state depends on their time-dependent 
behaviours. For thermalization, the $\q_i$'s must behave in such a way 
such that $x_i \ra \infty$ as $\t \ra \infty$. That means they  
must grow less fast than $\t$. In \fref{gr:relax_t}, we show these
$\q_i$'s as a function of $\t$. Initially, $\q_i >\t$ for both quarks 
and gluons, and $\qq$ starts off very large (see Table 1) but drops 
extremely rapidly back down to within hadronic timescales. The 
subsequent expected increase in time \cite{baymetal1,baymetal2,heis} 
is sufficiently slow for $\t$ to get past $\qg$ and $\qq$ 
at LHC, \fref{gr:relax_t} (a) and (b) but at RHIC, 
\fref{gr:relax_t} (b'), $\qq$ is still too large for $\t$ to overtake 
it before the temperature reaches $200$ MeV. Nevertheless, the 
$\t$-dependence is slow enough that $x_i$ should go to infinity 
as $\t \ra \infty$. 

\begin{figure}
\centerline{
\hbox{
{\psfig{figure=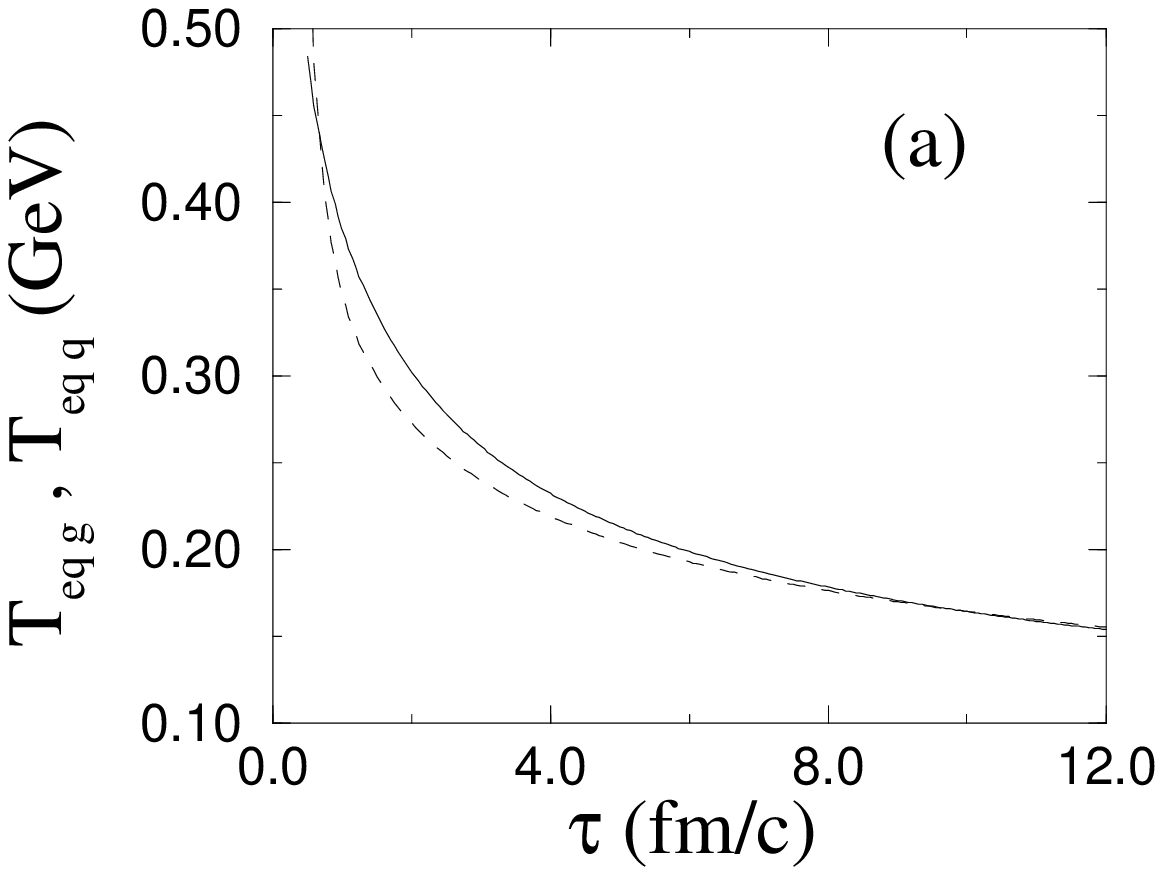,width=3.2in}} \ \
{\psfig{figure=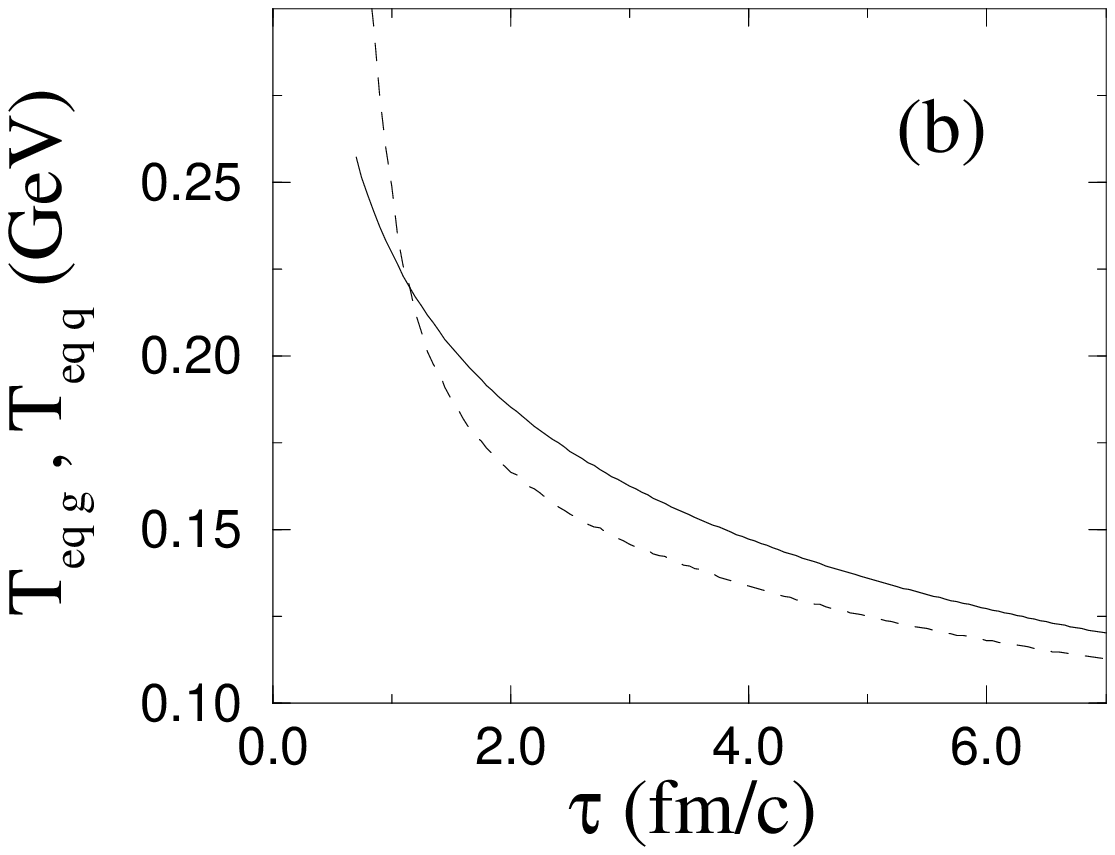,width=3.2in}}
}}
\caption{The time development of the equilibrium target 
gluon (solid line) and quark (dashed line) temperatures 
$\Tg$ and $\Tq$ respectively at (a) LHC and (b) RHIC. 
They should converge in an oscillating fashion at large 
$\t$ in order for the system to equilibrate as one towards 
a single temperature. The convergence is less good at RHIC 
than at LHC.}
\label{gr:T_eq}
\end{figure}

We have mentioned in Sect. \ref{sec:relax}, for the system to
equilibrate as one, the target equilibrium temperatures 
$\Tg$ and $\Tq$ and also $\qg$ and $\qq$ must approach each other
at large $\t$. We strongly suspect that the convergence of the 
temperatures will proceed in an oscillating fashion where the two 
curves intersect each other several times before the final convergence
at very large $\t$. We can see this in \fref{gr:T_eq} (a) and (b). 
At LHC, the initial condition is more favourable for equilibration 
and so $\Tg$ intersects $\Tq$ twice already. This is not so at RHIC. 
In fact, all indications point to the fact that a plasma created 
at LHC will equilibrate better than one created at RHIC. By letting
the plasma to continue its evolution and ignoring the deconfinement
phase transition, we have seen that the collective variables 
like the gluon and quark energy densities, gluon entropy
density etc. do show tendency to pass from below to above or vice
versa, the corresponding equilibrium target values i.e. 
tendency to overshoot the equilibrium values and hence oscillation.
As to the convergence of $\q_i$'s, it is not so clear 
in \fref{gr:relax_t} (c) and (c'), especially at RHIC in 
\fref{gr:relax_t} (c'). $\qq$ is much too large in comparison 
with $\qg$ for any clear sign of convergence within the 
time available. On the other hand, at LHC, although
there is still a large gap between the magnitudes, there is
a clear tendency that the rate of increase of $\qq$ with $\t$ is
slowing down in \fref{gr:relax_t} (b) while $\qg$ still increases at 
approximately the same rate. It is simply too early for the system 
to equilibrate as one. Even near the end, the quarks and gluons
can only be considered as two linked subsystems approaching
equilibrium at very different rates. Hence we have a two-stage
equilibration. 

Having shown that interactions can indeed dominate over the 
one-dimensional expansion of the parton gas in the central region
of relativistic heavy ion collisions and hence bring the  
plasma into equilibrium. We can now look at the individual processes 
and compare their relative importance. These are the processes
\etrref{eq:ggi}{eq:gqi}{eq:qqi}. We have labelled their contributions
to the gluon and quark collision entropy rate $ds_g/d\t$ and 
$ds_q/d\t$ by $d s_{gi}/d\t$, $i=1,\dots, 4$ and $d s_{qi}/d\t$, 
$i=1,\dots,3$ in the order that they appear in 
\etrref{eq:ggi}{eq:gqi}{eq:qqi}. Processes that give the same rate 
due to quark-antiquark symmetry are considered as the same process. 
Hence $gq \llra gq$ and $g\bar q \llra g\bar q$ give identical 
contribution to gluon and quark collision entropy density rate as 
$ds_{g4}/d\t$ and $ds_{q2}/d\t$ respectively. Also we have 
combined fermion elastic scattering processes as one rate
$ds_{q3}/d\t$ for convenience. These are shown in 
\fref{gr:s_g-lhc}, \fref{gr:s_g-rhic}, \fref{gr:s_q-lhc} and 
\fref{gr:s_q-rhic}. The elastic processes have a characteristic 
shape, i.e. an initial rapid rise to a peak at maximum anisotropy 
before returning to zero progressively. The sharper the peak, 
the quicker the kinetic equilibration (compare \fref{gr:s_g-lhc} 
and \fref{gr:s_g-rhic} (b), (d) and \fref{gr:s_q-lhc} and 
\fref{gr:s_q-rhic} (b), (c) and \fref{gr:press}). Note the negative
rate of \fref{gr:s_g-lhc} and \fref{gr:s_g-rhic} (c) which 
is because there are net quark-antiquark pair creations from 
gluon-gluon annihilations and entropy decreases with the 
number of gluons as already mentioned in the previous paragraphs. 
We compare the different processes by plotting the ratio of the 
magnitude of each contribution to that of gluon multiplication  
for gluons in \fref{gr:s_g-lhc} and \fref{gr:s_g-rhic} (e) and the 
ratio of each rate to that of quark-antiquark creation for quarks 
(antiquarks) in \fref{gr:s_q-lhc} and \fref{gr:s_q-rhic} (d). 
In the (e) figures, gluon multiplication clearly dominates initially 
at $\t \lsim 2$ fm/c at LHC and $\t \lsim 4$ fm/c at RHIC since all 
three ratios in each plot are less than 1. After these times, 
$q\bar q$ creation becomes dominant (thick solid line) and rises 
to several times larger than gluon multiplication. The $gg$ elastic 
scattering, on the other hand, tends to maintain a small, nearly 
constant ratio with gluon multiplication (solid line), which 
supports the claim made in \cite{wong}. That is, in a pure gluon
plasma, gluon multiplication dominates over $gg$ elastic
scattering in driving the plasma towards equilibrium. This 
remains the case even when $l_g \sim 0.93$
which shows that this dominance is not sensitive to the
value of $l_g$. The remaining ratio of quark-gluon scattering
to gluon multiplication continues to rise but not as rapidly
as the first ratio. For quark entropy, \fref{gr:s_q-lhc} and 
\fref{gr:s_q-rhic} (d), both ratios of quark-gluon 
scattering (solid line) and fermion-fermion scatterings 
(dashed line) to $gg \llra q\bar q$ rate remain small 
during the time available although they are
both on the rise. So for gluons, gluon multiplication
dominates initially but is later overtaken by $gg\llra q\bar q$ 
which continues to dominate over other elastic processes.
For quarks (antiquarks), this same process dominates during
the lifetime of the plasma. 

These behaviours can be understood in the following way.
Gluon branching dominates initially over any other processes
so long as gluons are not near equilibrium. Once they 
approach saturation (the $l_g$ estimates slow down their
approach towards $1.0$ in \fref{gr:T&l} (a) and (b) at about
the times mentioned above), gluon-gluon annihilation to 
quark-antiquark takes over as the dominant one because 
the fermions are still far from full equilibration. 
Because of the latter reason, the other ratios involving
quark or antiquark to gluon branching continue to rise.

So contrary to common assumption, inelastic processes are 
dominant in equilibration. This should have consequences in 
the perturbative calculations of transport coefficients or 
relaxation times \cite{baymetal1,baymetal2,heis} of system
that are not subjected to external forces. These calculations 
are based essentially, up to the present, on elastic binary 
interactions. As we have seen, they are not the dominant 
processes in equilibration. 

To the surprising result of gluon multiplication dominates
over elastic gluon-gluon scattering, we provide the following
explanation. If one only looks at the scattering cross-sections,
it is indeed true that gluon-gluon scattering has a larger
value and gluon multiplication processes are down by $\a_s$
for each extra gluon produced. The $(n-2)$ extra gluon production
cross-section can be expressed in terms of the elastic scattering
cross-section as \cite{gold&rosen,shury&xion}, in the double 
logarithmic approximation,
\be \s_{gg \ra (n-2) g} \propto \s_{gg \ra gg} \;
    [\a_s \ln^2 (s/s_{cut})]^{n-4} 
\ee
where $s_{cut}$ is the cutoff for the mininum binary invariant
$(p_i+p_j)^2 > s_{cut}$ of the 4-momenta of each gluon pair.
In the present problem, $s_{cut}=m^2_D$, the double logarithm 
is not large and certainly does not compensate for the smallness 
of $\a_s$. However, as we have mentioned at the beginning, 
the collision term on the r.h.s. of \eref{eq:baymeq} 
consists of the sum of the differences of the reactions in 
a QCD medium going forward and backward, so a large 
cross-section does not automatically imply dominance of 
the corresponding process in the approach to equilibrium. 

\noindent
Similarly, $gg \llra q\bar q$ is not that different from 
$gq \llra gq$ or $g\bar q \llra g\bar q$ because the
two matrix elements are related simply by a swapping
of the Mandelstam variables. So why should the first
dominates over the second? Except the different ways that 
the infrared divergences are cut off in the processes,
the main reason is $gg \lra q\bar q$ dominates over 
the backward reaction $q\bar q \lra gg$ due to the
simple fact that there are less fermions than gluons
present in the plasma. An extreme example of this 
phenomenon would be the forward and backward reaction 
balance out each other for all the elastic interactions as 
in a kinetically equilibrated plasma when only inelastic 
processes remain in the collision terms. In this extreme,
all the ratios of elastic to inelastic collision entropy 
rate vanish. 

We can now return to the question of whether other inelastic
processes such as $gg \llra q\bar qg$, $gq \llra gqg$,
$g\bar q \llra g\bar qg$, $gq \llra qq\bar q$, 
$g\bar q \llra q\bar q\bar q$, $qq \llra qqg$ etc. should be 
included. Although they are non-leading compared to 
$gg \llra ggg$ and $gg \llra q\bar q$ due to colour, they
should be significant when one sizes them with the elastic 
processes in view of the cancellation between the forward and 
backward reactions. In \cite{wong}, the question of the 
dominance of inelastic over elastic processes was raised. 
Here it is sufficient to include the two leading inelastic 
processes to show this explicitly. Had one included these other 
processes, then equilibration should be faster and one could 
end up with a more reasonable quark-antiquark content in 
the plasma. However, we are doubtful that the equilibration 
time can be reduced dramatically from what we have shown here. 

As we argued in \cite{wong}, it is hard to perturb a parton
system from thermal equilibrium without doing so chemically.
Therefore inelastic processes are always active in the 
approach to equilibrium whereas the same is not true for
elastic processes. From our figures, it can be seen
that inelastic processes are not there only for chemical
equilibration or for minor contributions to thermalization
as is commonly assumed due to their possible higher powers 
in $\a_s$, they contribute even more significantly to
equilibration than elastic processes. Changing the initial
conditions will only vary the dominancy but not remove
the dominance. 

Before closing, we would like to point out some differences
of our results with that of PCM. In PCM, there appears to be 
no early momentaneous isotropic particle momentum distribution
in either S+S or Au+Au collisions. The first time that  
there is approximate isotropy, it is already thermalization
according to \cite{geig}. It was claimed that there was
no further significant change in the total momentum
distribution after $\t=2.4$ fm/c for Au+Au collision at
RHIC. We assume that they mean the shape of the distribution
with the exception of the slope which should continue to change 
due to cooling. However, when the total distribution is broken
down into that of the parton components, the approximate
isotropy or thermalization becomes less obvious. We have
shown that thermalization in the strict sense is slow and 
isotropy of gluon momentum distribution can be argued to be 
approximate but that of the fermions is not so good. 

As to chemical equilibration, PCM shows little chance of
that for the fermions. The corresponding fugacity estimates
are approaching the ``wrong direction'' with increasing time.
This is due to a net outflow of particles from the defined
central region. The net flux of outgoing particles is 
arguably more important for fermions than for gluons because
the formers have a larger mean free path. The result is the
gluon (fermion) fraction of the particle composition rises
(drops) with increasing time. Therefore even if there is no 
phase transition and the parton plasma is allowed to continue 
its one-dimensional expansion indefinitely, chemical 
equilibration will never be achieved.
Then according to PCM, the expansion is slow enough for 
kinetic equilibration for all particle species but too fast
for chemical equilibration of the quarks and antiquarks.
The boundary effect is too important and is affecting 
equilibration. In our case, this effect is not incorporated. 
Although equilibration is slow, full equilibration will be 
reached given sufficient time. 

We find it surprising that although the gluon fugacity 
estimate in PCM \cite{geig&kap} overshoots and stays 
above or at $1.0$ nearly all the time except at 
the beginning, $R_g$ is still positive or an order
of magnitude larger than $R_q+R_{\bar q}$ when the fugacities
of the latter are well below $1.0$ and decreasing. One would 
expect rather gluon absorption or conversion into quark-antiquark 
should take a significant toll on the gluon production so that
there should be a diminution of gluons. At least, this 
should be the case when local kinetic equilibrium has been 
or nearly been reached which PCM claimed to be so at the 
end of the program run but this is not the case in the 
plot of the production rate of the different particle species!
This is counter-intuitive and opposite to what we have shown. 

To conclude, we have shown that inelastic processes dominate
in the approach towards equilibrium. In particular, gluon 
branching is most important. Gluon-gluon annihilation into 
quark-antiquark becomes more important only when the gluons 
are near saturation and equilibrium. The lower power in
$\a_s$ of the gluon-gluon elastic scattering as compared to the 
inelastic gluon emission process is more than compensated for by 
the cancellation of the reaction going forward and backward. 
The recovery of isotropy in momentum distribution is slow
and so is chemical equilibration. The latter is partly due
to the small initial fugacities that we used. As an intrinsic
feature of perturbative QCD, the quarks and antiquarks 
are lagging behind the gluons in equilibration and hence a
two-stage equilibration scenario.

\section*{Acknowledgements}

The author would like to thank M. Fontannaz, D. Schiff and 
everyone at Orsay for kind hospitality during his stay 
there, R.D. Pisarski and A.K. Rebhan for raising interesting 
questions. Thanks also go to R. Baier and everyone at 
Bielefeld for hospitality during the author's short stay
there where this work is completed. The author acknowledges 
financial support from the Leverhulme Trust.

\end{document}